%% file: iclr2026_conference.tex
\documentclass{article} % For LaTeX2e
\usepackage{iclr2026_conference,times}

\input{math_commands.tex}

\usepackage{times}
\usepackage{graphicx}
\usepackage{booktabs}
\usepackage{enumitem}
\usepackage{amsmath}
\usepackage{amssymb}
\usepackage{xcolor}
\usepackage{caption}
\usepackage{hyperref}
\usepackage{svg}
\usepackage{multirow}
\usepackage{bm}
\usepackage{placeins}
\usepackage{float}
\usepackage{microtype} 
\usepackage{subcaption}
\usepackage{url}

\definecolor{bestgreen}{RGB}{34,139,34}

\title{Multimodal Data Curation Through Ranked Retrieval}

\author{Pratyush Muthukumar, Harshil Kotamreddy, Sarah Amiraslani, Tomo Kanazawa, \\
\textbf{Ramani Akkati, Shaan Jain, \& Andrew Mathau} \\
NVIDIA \\
\texttt{\{pmuthukumar, hkotamreddy, samiraslani, tkanazawa,} \\
\texttt{rakkati, shaanj, amathau\}@nvidia.com}
}

\iclrfinalcopy % Uncomment for camera-ready version, but NOT for submission.
\begin{document}

\maketitle

\begin{abstract}
Shared embedding spaces are widely used for multimodal search and data curation. In practice, two problems often limit how well this works. First, embeddings can reflect modality more than meaning, so examples cluster by input type even when the underlying content matches. Second, the paired supervision used to train these spaces is often noisy. 
% Captions may omit key details, transcripts can include irrelevant speech, and annotations may be only loosely grounded to the signal. 
When we blend many heterogeneous, human-labeled datasets, these issues reinforce each other and degrade cross-modal retrieval. We present a framework that improves alignment by acting on both the training pairs and the embedding model. Symmetric Nucleus Subsampling (SNS) refines training pairs by trimming raw inputs and annotations to the portions that best support each other. Expert Embedding Engine (EEE) combines complementary embedding experts using a learned projection network, together with a bias-aware objective that reduces modality-driven separation in the embedding space. We demonstrate that this approach collapses the modality gap by over $90$\% on average vs base embedding experts and is a strong data curator, with datablends from our method outperforming stratified sampling and traditional curation baselines in downstream model performance. 
\end{abstract}

\section{Introduction}

Consider a user searching a large archive that mixes images, clips, transcripts, and audio recordings. An ideal state would be a single interface where a short query reliably surfaces the right content, regardless of whether the answer lives in a frame, a caption, or a few seconds of sound. This expectation is a big reason shared-embedding approaches (popularized by contrastive vision–language training) have become the default backbone for multimodal search, recommendation, and dataset analytics. Increasingly, the same similarity-search loop is also used upstream to curate the training mixtures that produce these models.

\begin{figure}[H]
    \centering
    \includegraphics[width=0.75\linewidth]{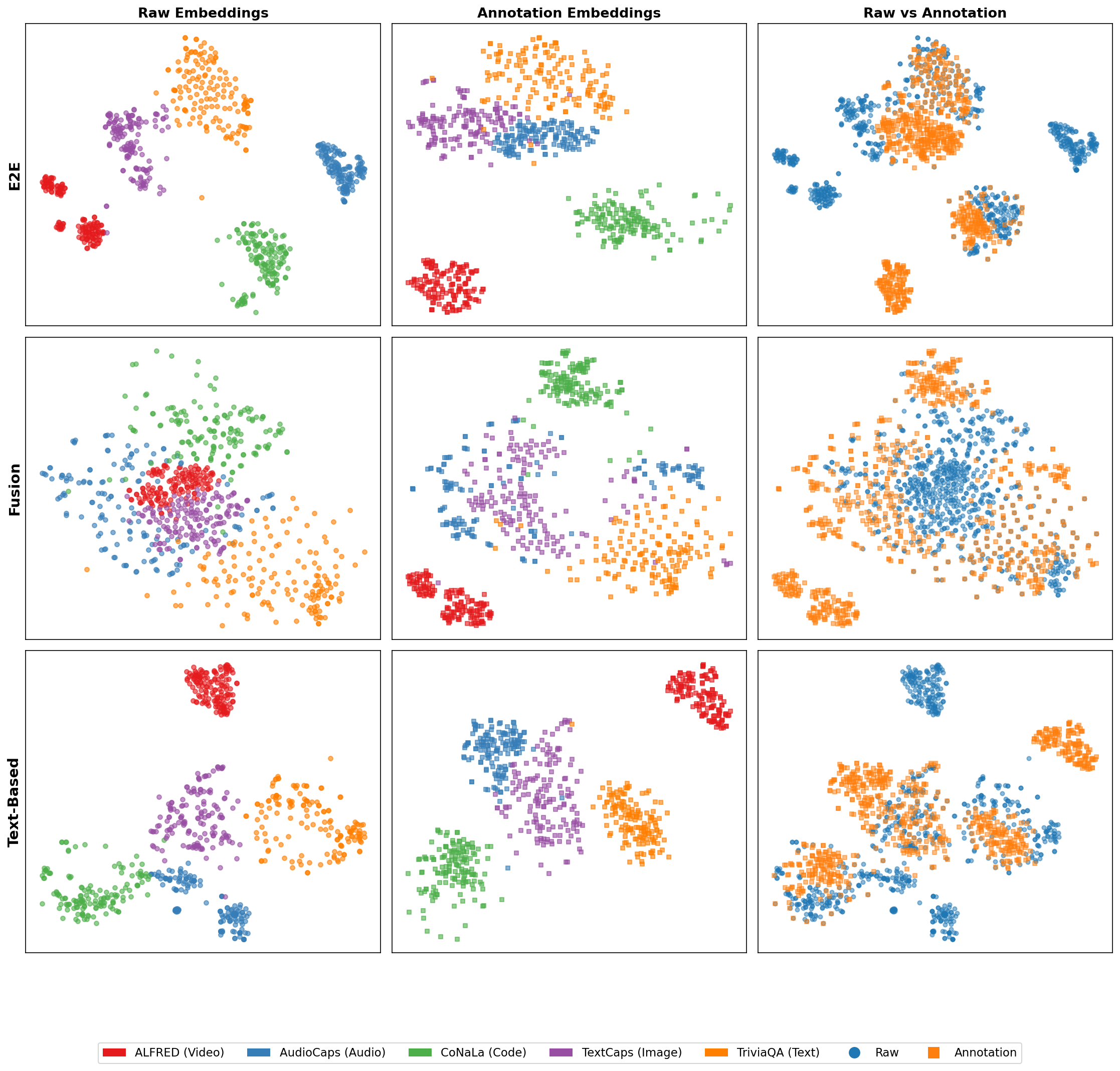}
    \caption{A 2D t-SNE visualization of paired data embeddings by common multimodal embedding expert implementations (text-based, fusion, end-to-end), illustrating modality-dependent clustering.}
    \label{fig:modality_gap}
\end{figure}

This shift toward embedding-first systems also mirrors a broader trend in modern model training. For large language models and multimodal models alike, data quality and data composition increasingly determine what the model learns. When we scale up training, we rarely rely on a single clean dataset. We blend many sources, each with its own annotation style, noise patterns, and coverage. In that setting, curation becomes a first-class problem: which examples should we include, which should we downweight, and how do we combine datasets without washing out rare but valuable signal? Search-based retrieval is a practical way to curate across many corpora because it can surface semantically related examples, make mixtures easy to audit, and let us iterate quickly. At the same time, this raises the stakes for the embedding space itself, since retrieval will only be reliable if distance actually tracks meaning. That dependence on the embedding space hides a fragile assumption: that distance primarily reflects semantic similarity. In real deployments, the geometry often reflects something else just as strongly - modality identity. Text tends to cluster with text, images with images, and audio with audio, even when examples are describing the same underlying concept (Figure \ref{fig:modality_gap}). This directly affects cross-modal retrieval, where nearest neighbors may share modality rather than meaning. It also complicates fusion strategies that rely on consistent geometry and reduces the reliability of downstream embedding workflows like clustering, deduplication, and active data selection.

A second, independent issue is that the training signal used to create these embedding spaces is rarely clean. Paired datasets often exhibit raw–annotation misalignment: captions omit important objects or actions, transcripts include irrelevant context, and annotations can refer to content that is off-screen or inaudible. When supervision is only partially grounded, the model is effectively asked to align a raw sample with an annotation that shares limited information. In response, training can reward shortcuts - features that correlate with the dataset or modality rather than the described event - which in turn reinforces modality bias instead of correcting it.

Most existing solutions tackle one side of this problem at a time. Representation-level methods aim to reduce modality separation through calibration or alignment strategies, but typically accept the supervision signal as given \citep{long2024llms}. Data-centric filtering and reweighting strategies try to improve pair quality, but often assume a fixed encoder family and do not explicitly address systematic modality-driven geometry \citep{zha2023datacentric}. In settings where we want to merge datasets and learn a unified embedding space that supports robust search and analysis, these issues are difficult to decouple: noisy pairs make geometry worse, and biased geometry makes it harder to detect which pairs are noisy.

We therefore take a coupled approach that intervenes on both the examples and the embedding model. On the data side, we introduce Symmetric Nucleus Subsampling (SNS). Rather than treating each paired sample as atomic, SNS identifies high-information nuclei on both sides of the pair by selecting the most relevant portions of the raw input with respect to the annotation, and the most grounded portions of the annotation with respect to the raw input, then retains content that remains consistent under a symmetric similarity check. The goal is not to “clean” the dataset into a different distribution, but to make each training pair carry a clearer alignment signal while preserving a safe fallback to the original pair when trimming does not improve alignment.

On the model side, we introduce an Expert Embedding Engine (EEE), a mixture-of-experts embedding architecture that combines embedding spaces through a learned projection mechanism. The experts are designed to capture different strengths (e.g., text-centric representations which are useful when strong text encoders are available), fusion-based representations (useful when jointly modeling modalities), and modality-unified encoders so that the system can adapt to the input and annotation characteristics of each example. Training includes a bias-aware objective that discourages modality-driven separation. 

%Because non-text modalities often benefit from being summarized in language (especially when using text-centric experts), we incorporate structured modality-to-text prompting for the text-centric expert to produce stable descriptive signals that integrate naturally into the embedding pipeline.

We study this framework in practical retrieval and clustering settings over image–text, audio-text, video-text, and text–text collections. Our evaluation focuses not only on downstream metrics but also on embedding-space behavior, including modality-wise separation and clustering structure, and on how data-side subsampling and model-side projection interact.

Our contributions are as follows:

\begin{itemize}
    \item We analyze how modality bias and raw–annotation misalignment jointly affect multimodal embedding quality, particularly in mixed, human-labeled corpora.
    \item We propose SNS, a symmetric, annotation-aware subsampling method that improves pair consistency by trimming irrelevant content on either side of a multimodal pair.
    \item We propose EEE, a mixture-of-experts embedding engine with a learned projection layer applying a bias-aware objective to reduce modality-driven separation while preserving semantic neighborhood structure.
    \item We design a robust evaluation study using a human-selected task set for balanced comparisons of downstream model performance impact by datablends curated through variants of our approach and industry standard baselines.
\end{itemize}

\section{Preliminaries and Background}

\subsection{Paired multi-modal datasets}
We define paired multi-modal datasets as a collection of paired samples $\{(x_i, y_i)\}_{i=1}^{n}$, where $x_i \in \mathcal{X}$ is a \emph{raw data} view in one high-bandwidth modality (e.g., image, video, long text) and  $y_i \in \mathcal{Y}$ is a supervisory description and/or label \emph{annotation data} view in another modality (typically text). We embed both raw and annotation data as $\ell_2$-normalized vectors for clustering, retrieval, and further manipulations.

\subsection{Contrastive Learning (CLIP) and Mutual Information}
Central to modern retrieval and transfer learning is the concept of dual-encoder setups, wherein encoders $f_x$, $f_y$ map paired data $(x_i, y_i)$ to embeddings $z\{x_i\}$, $z\{y_i\}$ in a shared embedding space with a training objective to increase similarity of matched pairs $(x_i^+, y_i^+)$ and decreasing similarity of \emph{in-batch negative} mismatched pairs $(x_i^-, y_i^-)$ in shared space; a temperature parameter scales logits/similarities in this contrastive objective (\cite{radford2021learning}. Key to this formulation of contrastive learning is the assumption that paired samples $(x_i, y_i)$ share more semantic content than mismatched pairs, so ‘positive’ (semantic) similarities should be systematically higher than ‘negative’ similarities (shared modality).

Conceptually, contrastive objectives are often framed as an attempt to maximize \emph{mutual information} (MI) between paired data views \citep{oord2019representationlearningcontrastivepredictive}. This relies on certain sampling assumptions that can degrade in high-dimensional spaces/with large MI \citep{poole2019variational}, so we treat MI as an intuition and rely on similarity-based proxies rather than attempting to estimate it directly.

\subsection{Failure modes: Modality gap and data misalignment}
As alluded to above, multi-modal embedding models face a fundamental challenge: individual embedding functions can develop geometric separation, in which embeddings cluster by modality identity rather than by shared semantic content. This modality gap can harm semantic classification and retrieval, even when paired samples have high semantic similarity \citep{liang2022mind}.

In addition to the modality gap, paired data also commonly contain other types of misalignment:
\begin{itemize}
    \item \textbf{Extraneous/off-target annotation:} annotation contains irrelevant context not observable in or supported by the raw data view (e.g., off-topic tangents in image description).
    \item \textbf{Incomplete/under-descriptive annotation:} annotation incompletely spans salient raw data content; this can be spatial or temporal (e.g., in a 30-minute automotive dash camera video, only describe street signs or only describe the last 5 minutes of activity, respectively)
\end{itemize}

Misaligned sample pairs violate the assumption that matched pairs share key semantic content, thereby amplifying geometric artifacts such as modality gaps and further complicating cross-modal retrieval.

\section{Related Work}

As a result, data selection and filtering are widely recognized as critical in large-scale representation learning, particularly for contrastive pretraining where web-scale corpora contain noise, duplication, and weakly aligned pairs. Recent work studies practical filtering strategies and training refinements that improve contrastive vision language pretraining under noisy data \citep{radenovic2023diht}. Dataset curation benchmarks and systematic comparisons of filtering, reweighting, and sampling strategies for training multimodal models highlight that quality and mixture choices can dominate downstream outcomes \citep{gadre2023datacomp}. Additional discussion of related work can be found in Appendix \ref{ref:related}.

% From a different angle, interpretability work such as Sufficient Input Subsets (SIS) identifies minimal subsets of an input that preserve a model's prediction, exposing spurious shortcuts and offering instance level explanations \citep{carter2019sis}. While SIS is not a data selection method, it motivates the broader notion that subset structure can reveal which parts of an example carry signal versus noise.

% \paragraph{Relation to our work.}
% Our Symmetric Nucleus Subsampling, SNS, is closest to the data centric strand, but differs from typical filtering approaches in two ways. First, SNS is annotation aware in that it operates over paired multimodal inputs and their associated labels or tags, not just raw pairs treated as atomic units. Second, SNS is symmetric. It selects or downweights examples by gating on informativeness in both directions, from raw modality to annotation and from annotation to raw modality. This is especially relevant when merging many labeled datasets with inconsistent label structures and variable annotation noise. Unlike SIS, which targets faithful explanations, SNS is designed as a training data resampling mechanism aimed at improving retrieval behavior and reducing modality gaps in a combined training pool.

\section{Methods}
To rank relevant data samples across several human-labeled datasets and modalities, we devise a three-step approach:

\begin{itemize}
    \item First, we use SNS to reduce misalignment within raw data and annotation pairs by removing irrelevant portions of the raw data and/or the annotation.
    \item Second, we use EEE to reduce modality-specific bias from any specific embedding model by using several embedding models with different multi-modal approaches.
    \item Finally, in order to rank embeddings generated by the EEE, we combine the embedding spaces of each of the experts using a projection network. 
\end{itemize}

While we note the use of specific models in the first two steps, these models can be switched out for any other model that performs the same function. We intend for this method to be a generalizable framework without restriction on specific model usage.

\subsection{Reducing Misalignment with Symmetric Nucleus Subsampling}
Paired multimodal datasets often have raw data that contain information that is not included in the annotation/label, and vice versa. More concretely, for a paired data sample $(x, y) \in \mathcal{X} \times \mathcal{Y}$, misalignment occurs when $I(x; y) \ll \min(H(x), H(y))$, i.e., the mutual information $I$ is much less than the entropy $H$ of either component. We quantify this using the \textit{information density}, $\phi$, of a paired data sample, which is defined as the ratio of mutual information to the total content size:

\begin{equation}
\phi(x, y) = \frac{I(x; y)}{|x| + |y|}.
\end{equation}

To increase $\phi$ for any given paired data sample, we propose extracting nuclei from both the raw data and the annotation. This is done through a forward extraction phase and a backward extraction phase. In forward extraction, we extract the most relevant part (nucleus) of the raw data based on the contents of the annotation. In backward extraction, we do the opposite, i.e., we extract the nucleus from the annotation based on the contents of the raw data.  

\begin{figure}[t]
    \centering
    \includegraphics[width=0.8\linewidth]{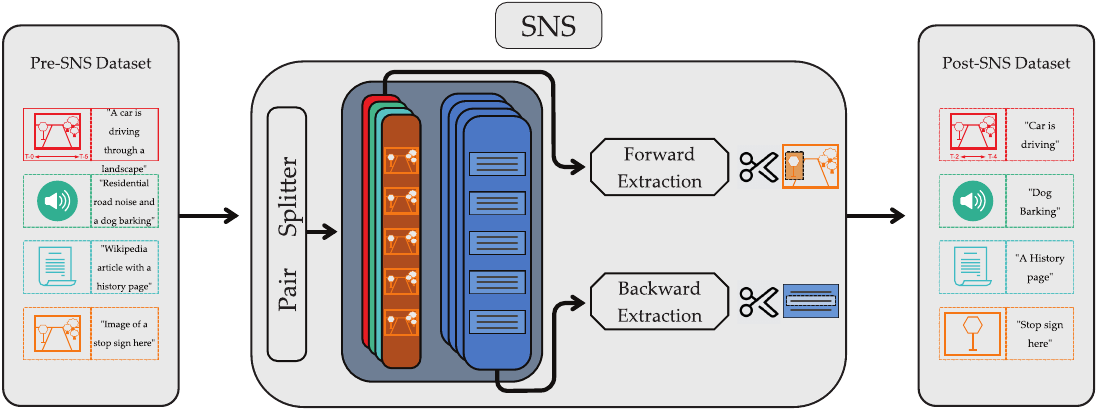}
    \caption{Symmetric Nucleus Subsampler component overview, including Forward Extraction and Backward Extraction modules for effective reduction of misalignment in paired samples}
    \label{fig:sns}
\end{figure}

In this work, we use the result of forward extraction (the nucleus of the raw data) when performing backward extraction to maintain semantic consistency between the raw data and the annotation. The overall architecture of the Symmetric Nucleus Subsampler component is visualized in Figure \ref{fig:sns}.

\subsubsection{Forward Extraction}
In the forward extraction phase, we address data-to-annotation misalignment by identifying portions of raw data that maximally explain the annotation content. Let $\mathcal{N}_{\alpha} : \mathcal{X} \times \mathcal{Y} \rightarrow \mathcal{X}$ be the forward extraction function. $\mathcal{N}_{\alpha}$ outputs a subset of the raw data $\tilde{x}$ such that $I(\tilde{x}; y) > \rho \cdot I(x;y)$, where $\rho$ is the mutual information ratio. If no such subset exists, $\mathcal{N}_{\alpha}$ simply outputs the original raw data. 

In practice, $\mathcal{N}_\alpha$ is implemented using several modality-specific segmentation methods and mutual information is approximated using embedding similarity. We use Omni-Embed-Nemotron-3B \citep{xu2025omniembednemotronunifiedmultimodalretrieval} to compute embeddings across all modalities. While this still induces modality-specific bias, it serves as a noisy proxy to mutual information (see Figure \ref{fig:pairwise_sim}). The segmentation methods for each modality are as follows:

\textbf{Text:} We use Omni-Embed-Nemotron-3B \citep{xu2025omniembednemotronunifiedmultimodalretrieval} to embed the annotation and each sentence of the corresponding raw data. We then calculate the cosine similarity between each sentence and the annotation and keep a sentence if the similarity is higher than $\tau_{\alpha}$. 

\textbf{Image:} We use the Grounding-DINO \citep{liu2024groundingdinomarryingdino} image segmentation model to create bounding boxes for the most relevant parts of the image based on the annotation. We keep the bounding boxes with confidence scores higher than $\tau_{\alpha}$. If there are multiple bounding boxes, we get the minimum spanning bounding box. We then extract the part of the image within the bounding box. 

\textbf{Video and Audio:} We use moment detection models (CG-DETR \cite{moon2023correlation} for video and AM-DETR \cite{munakata2025language} for audio) to extract the most relevant time span within the video/audio data to the annotation. We keep the time spans with confidence scores higher than $\tau_{\alpha}$. If there are multiple time spans within the video that are relevant to the annotation, we remove all data between the extracted time spans and splice together the relevant portions.

In practice, $\tau_\alpha$ can have different values for each modality; however, we present it as a single value for simplicity. Once a subset $\tilde{x}$ of the raw data $x$ is extracted, we check that $I(\tilde{x}; y) > \rho \cdot I(x;y)$ before replacing $x$ with $\tilde{x}$.

\subsubsection{Backward Extraction} \label{sec:backward_extraction}
In the backward extraction phase, we address annotation-to-data misalignment by identifying portions of the annotation that maximally explain the content of the raw data. Let $\mathcal{N}_\beta : \mathcal{X} \times \mathcal{Y} \rightarrow \mathcal{Y}$ be the backward extraction function. $\mathcal{N}_\beta$ outputs a subset of the annotation $\tilde{y}$ such that $I(x; \tilde{y}) > \rho \cdot I(x; y)$ where $\rho$ is the mutual information ratio. If no such subset exists, $\mathcal{N}_\beta$ simply outputs the original annotation.
In practice, we implement $\mathcal{N}_\beta$ using several modality-to-text models by first converting the raw data to a description and then using embedding similarity to extract the most relevant sentences from the annotation. The models we use to describe the raw data are as follows:

\textbf{Text:} We simply use the text raw data with no further processing.

\textbf{Image:} We use the Nemotron-Nano-12B-v2-VL model \citep{nvidia2025nvidianemotronnanov2} to generate a description of the image.

\textbf{Video:} We use the Cosmos-Reason2-2B model \citep{nvidia2025cosmosreason2} to generate a description of the video. \looseness=-1

\textbf{Audio:} We use the Phi-4-multimodal-instruct model \citep{phi4mini2025}to generate a description of the audio.

To compute embeddings, we once again use Omni-Embed-Nemotron-3B \citep{xu2025omniembednemotronunifiedmultimodalretrieval}. We keep sentences from the annotation if their cosine similarity with the description generated is greater than some threshold $\tau_\beta$. Finally, we replace the original annotation $y$ with the extracted annotation $\tilde{y}$ if $I(x; \tilde{y}) > \rho \cdot I(x; y)$. We ablate values of $\tau_\alpha$, $\tau_\beta$, and $\rho$ in Appendix \ref{ref:sns_ablations}.

\subsection{Reducing Modality-Specific Bias with the Expert Embedding Engine}

Once data pairs pass through the SNS process, they are more closely aligned. We now discuss embedding each of these data pairs in order to retrieve the most relevant data pairs to some query $q$. While we can simply use a single multi-modal embedding model, we find that these models tend to have modality gaps, i.e., data points of a certain modality tend to be close to each other (Figure \ref{fig:modality_gap}).

To address modality gaps, we propose using several embedding models (experts) that compute embeddings differently. Each expert will tend to exhibit a different modality bias, giving us more information about the semantic meaning of each data pair. Specifically, we use the following embedding models:

\textbf{End-to-End Expert:} We use the Omni-Embed-Nemotron-3B multimodal embedding model \citep{xu2025omniembednemotronunifiedmultimodalretrieval}, which uses an end-to-end architecture to compute embeddings for all modalities.

\textbf{Fusion Expert:} We use the ImageBind fusion model \citep{girdhar2023imagebind}, which uses an embedding fusion architecture to combine several embedding spaces while using images as an anchor.

\textbf{Text Expert:} We first convert data of all modalities to text using the same modality-to-text models used in Section \ref{sec:backward_extraction}. Then, we use the Llama-Nemotron-Embed-1B text embedding model \citep{moreira2024nv} to embed the text descriptions.

However, we now run into a different problem: how do we rank multiple distinct embedding spaces?

\subsection{Combining Multiple Embedding Spaces With a Projection Network}

We train a lightweight neural network to adaptively combine embeddings from all three expert embedding engines into a unified representation. Without this network, each embedding space returns results separately at query time, making it difficult to rank relevance across different experts.

The network takes the concatenated embeddings from all $K$ experts as input (dimension $K \times d$) and outputs a single embedding vector of dimension $d$:
\begin{equation}
\mathbf{e}_\text{fused} = f_\theta([\mathbf{e}_1; \mathbf{e}_2; \ldots; \mathbf{e}_K])
\end{equation}
where $f_\theta$ is a multi-layer neural network and $[\cdot;\cdot]$ denotes concatenation.

\begin{figure}[ht]
    \centering
    \includegraphics[width=0.45\linewidth]{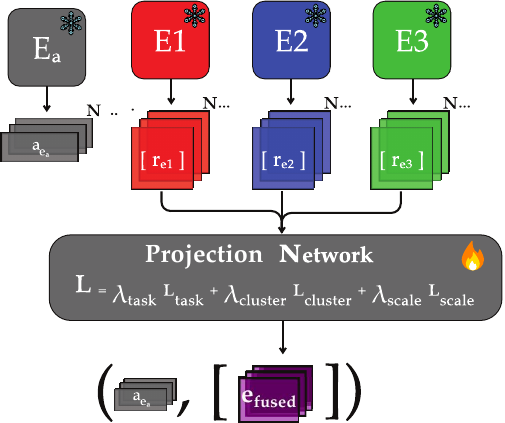}
    \caption{Expert Embedding Engine (EEE) \& Projection Network components. The learned embedding space of paired samples can be used to curate datablends using embedding similarity to the query vector.}

        \label{fig:eee}
\end{figure}

There is one exception to this fusion process: we need a grounded set of embeddings that act as anchors, of which the network will learn the geometry of the embedding space and guide the data sample embeddings to their respective locations accordingly. We treat the \emph{annotations} in the paired representation of data samples \textit{(annotation, raw data)} as these anchors. Thus, raw data samples of varied modalities are passed through the projection layer to arrive at $\mathbf{e}_\text{fused}$, but annotation embeddings are left unchanged. In practice, since our dataset consisted of all text annotations, we opted to simply use the embeddings generated from the text-based expert as these anchor embeddings, as depicted in Figure \ref{fig:eee}.

\subsubsection{Loss Function Formulation}

The projection network is trained with two objectives: (1) preserving semantic similarity between paired data and annotations, and (2) minimizing the modality gap in the fused embedding space.

The total loss has three terms:

\textbf{Task Loss:} An InfoNCE-style contrastive loss that encourages positive pairs (data and its annotation) to be close while pushing negative pairs apart:
\begin{equation}
L_\text{task} = -\log \frac{\exp(\text{sim}(\mathbf{x}, \mathbf{x}^+) / \tau)}{\sum_{\mathbf{x}_i \in \mathcal{B} \setminus \{\mathbf{x}\}} \exp(\text{sim}(\mathbf{x}, \mathbf{x}_i) / \tau)}
\end{equation}
where $\tau$ is the temperature parameter and the denominator sums over all in-batch samples except the anchor.

\textbf{Cluster Bias:} Penalizes the distance between each modality centroid and the overall centroid, encouraging different modalities to overlap in the embedding space:
\begin{equation}
L_\text{cluster} = \sum_{m \in \mathcal{M}} \| \boldsymbol{\mu}_m - \boldsymbol{\mu} \|_2^2
\end{equation}
where $\boldsymbol{\mu}_m$ is the mean embedding of modality $m$, and $\boldsymbol{\mu}$ is the overall mean across all modalities $\mathcal{M}$.

\textbf{Scale Bias:} Penalizes differences in cluster spread between each modality and the overall distribution:
\begin{equation}
L_\text{scale} = \sum_{m \in \mathcal{M}} \left| \sigma_m - \sigma \right|
\end{equation}
where $\sigma_m$ is the average distance of modality $m$'s embeddings from $\boldsymbol{\mu}_m$, and $\sigma$ is the corresponding overall spread.

The total loss is:
\begin{equation}
L_\text{total} = \lambda_\text{task} L_\text{task} + \lambda_\text{cluster} L_\text{cluster} + \lambda_\text{scale} L_\text{scale}
\end{equation}

We run a comprehensive ablation study to arrive at final $\lambda_\text{task}, \lambda_\text{cluster},$ and $\lambda_{scale}$ terms as well as projection network architecture, details described in Appendix \ref{ref:proj-ablations}.

Once embeddings are passed through the projection network, they are part of a unified embedding space and can now be used for retrieval. During query time, queries are passed through the text-based expert in a similar manner to the annotations used for grounding. 

\begin{figure}
    \centering
    \includegraphics[width=0.7\linewidth]{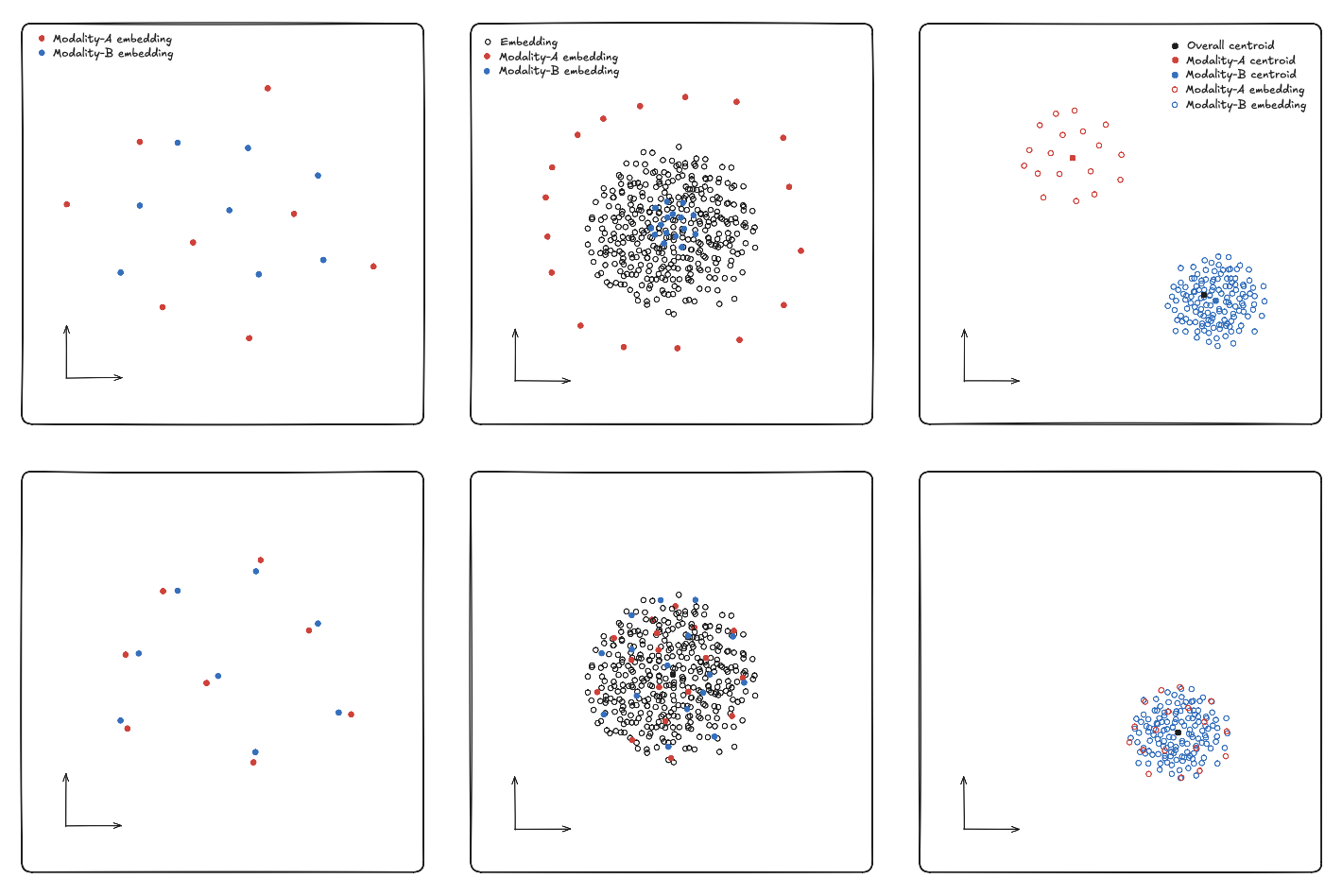}
    \caption{Effect of bias loss terms on modality gap reduction. Top row: Before training, embeddings from different modalities (A: red, B: blue) occupy distinct regions with separated centroids and varying cluster sizes. Bottom row: After training with cluster center bias ($L_{cluster}$) and scale bias ($L_{scale}$), modality embeddings overlap in a shared region, centroids align, and cluster spreads become uniform—enabling fair cross-modal retrieval. Appendix \ref{ref:proj-geometry} describes how this proposed effect is demonstrated in our empirical data.}
    \label{fig:loss terms}
\end{figure}

\section{Results}

\subsection{Datasets}

%% explain 5 pools, why we chose that data, what the modalities are

The candidate data pools comprises five human-annotated multimodal datasets spanning four modalities: \textbf{AudioCaps}~\citep{kim2019audiocaps} (audio captioning), \textbf{TextCaps}~\citep{sidorov2020textcaps} (image captioning), \textbf{ALFRED}~\citep{shridhar2020alfred} (video instruction following), \textbf{CoNaLa}~\citep{yin2018conala} (code captioning), and \textbf{TriviaQA}~\citep{joshi2017triviaqa} (text-text uestion answering). Each dataset stores paired \textit{(raw data, annotation)} entries, where the raw modality varies (audio, image, video, or text) and the annotation is fixed as natural-language text. 
% We chose these pools to ensure broad modality coverage---audio, image, video, and text---while providing human-verified alignment between raw content and its textual description.
From this combined pool of $10$k candidates ($2$k per pool), each curation strategy selects a fixed blend of $n{=}5{,}000$ samples for downstream fine-tuning.

\subsection{Downstream Evaluation Study}

%% explain what the downstream task is
%% details on training, models
Each curated blend of 5{,}000 samples is converted into an instruction-tuning dataset by wrapping every \textit{(raw data, annotation)} pair into a prompt--completion format. The model receives the raw media alongside a natural-language instruction (e.g., ``\textit{Describe the media content in detail}'') and is trained to generate the human-written annotation as the target response. This transforms the curation problem into a direct measure of multimodal instruction-following quality---better-curated blends should yield more informative supervision and, consequently, lower validation perplexity.

We fine-tune \textbf{Qwen2.5-Omni-3B}~\citep{xu2025qwen25omni}, a multimodal causal language model, using parameter-efficient LoRA adaptation~\citep{hu2022lora} on the curated blends. Each configuration is trained for 750 steps and replicated three times to compute 95\% confidence intervals. Validation perplexity serves as the primary metric, providing a task-agnostic signal of how well the fine-tuned model absorbs the curated data distribution without over-fitting to any single downstream benchmark.

Validation perplexity is computed on a static, human-curated held-out set of 500 pairs for all variants and baselines. To construct this set, we draw from an unseen partition of the same five data pools using a blind random shuffle, focusing on ``natural, real-world scenes containing objects, landscapes, subjects, or people'' -- which is the text query we use to curate blends for the curation task. Two of the authors of this study independently reviewed the shuffled candidates, and each selected 250 pairs they judged to best exemplify the query, yielding a combined evaluation set of 500 pairs (223 audio-text, 204 image-text, 53 video-text, 20 text-text).

We evaluate the following variants of our proposed architecture in this downstream evaluation study:
\begin{itemize}[nosep,leftmargin=*]
  \item \emph{EEE + Projection Only}: Expert Embedding Engine with learned projection; no SNS.
  \item \emph{SNS Forward + EEE + Projection}: Adds forward Symmetric Nucleus Subsampling, which filters raw data conditioned on the annotation.
  \item \emph{SNS Backward + EEE + Projection}: Uses backward SNS instead of forward SNS, extracting annotation-relevant substructures from the raw data.
  \item \emph{SNS Bidirectional + EEE + Projection}: Combines forward and backward SNS to capture both directions of the data--annotation relationship, along with EEE + projection network.
\end{itemize}

\begin{figure}[t]
    \centering
    \includegraphics[width=\linewidth]{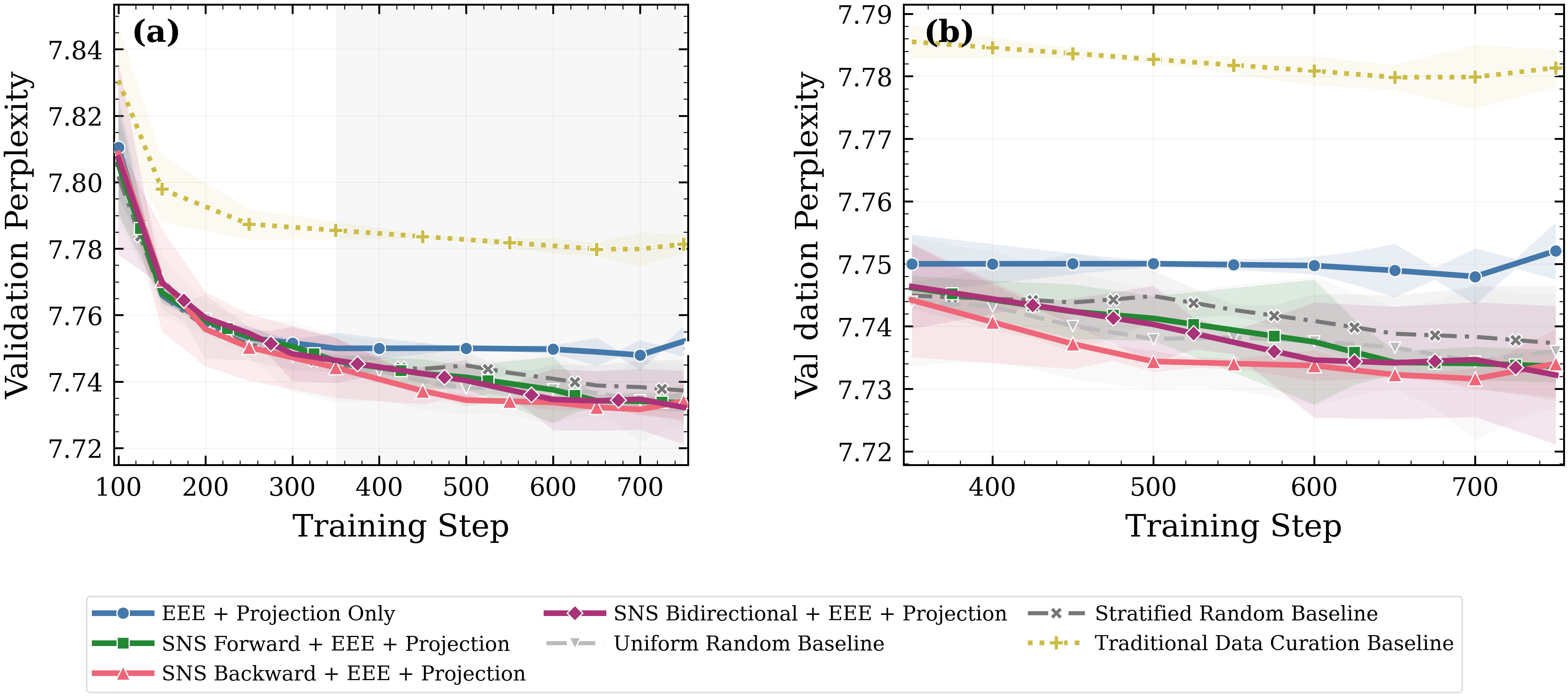}
    \caption{Validation perplexity (mean $\pm$ 95\% CI, $n{=}3$) across SNS \& EEE configuration variants vs baselines. \textbf{(a).} Full validation perplexity curves, \textbf{(b).} Last epoch validation perplexity curves.}
    \label{fig:val_perplexity}
\end{figure}

\begin{table}[b]
  \centering
  \caption{Validation perplexity (mean $\pm$ 95\% CI, $n{=}3$) at selected training steps. \textbf{Bold} indicates the lowest (best) perplexity per column. The \textit{Best PPL} column indicates the best perplexity over all steps.} \looseness=-1
  \label{tab:val_perplexity}
  \footnotesize
  \setlength{\tabcolsep}{4pt}
  \begin{tabular}{lcccc}
    \toprule
    \textbf{Method} & \textbf{Step 250} & \textbf{Step 500} & \textbf{Step 750} & \textbf{Best PPL} \\
    \midrule
    EEE + Projection Only & $7.753 \pm 0.004$ & $7.750 \pm 0.001$ & $7.752 \pm 0.005$ & $7.748 \pm 0.005$ \\
    SNS Forward + EEE + Projection & $7.754 \pm 0.001$ & $7.741 \pm 0.004$ & $7.734 \pm 0.003$ & $7.734 \pm 0.003$ \\
    SNS Backward + EEE + Projection & $\bm{7.750 \pm 0.010}$ & $\bm{7.734 \pm 0.002}$ & $7.734 \pm 0.005$ & $\bm{7.732 \pm 0.002}$ \\
    SNS Bidirectional + EEE + Projection & $7.755 \pm 0.001$ & $7.740 \pm 0.006$ & $\bm{7.732 \pm 0.011}$ & $\bm{7.732 \pm 0.011}$ \\
    \midrule
    Uniform Random Baseline & $7.753 \pm 0.006$ & $7.738 \pm 0.008$ & $7.736 \pm 0.009$ & $7.734 \pm 0.013$ \\
    Stratified Random Baseline & $7.751 \pm 0.005$ & $7.745 \pm 0.004$ & $7.737 \pm 0.009$ & $7.737 \pm 0.009$ \\
    Traditional Data Curation Baseline & $7.787 \pm 0.005$ & $7.783 \pm 0.001$ & $7.781 \pm 0.003$ & $7.780 \pm 0.002$ \\
    \bottomrule
  \end{tabular}
\end{table}

\subsection{Baselines}

%% what are good baselines in this case, why we couldn't apply prior works
\begin{itemize}
    \item \textbf{``Traditional'' Data Curation Pipeline}
To evaluate our method against common industry practices for curating datablends, we implement a baseline that combines heuristic filtering with semantic ranking. Further details on this approach are available in Appendix \ref{ref:baselines}.
    \item \textbf{Uniform Random Sampling.} We randomly sample $k=5{,}000$ pairs uniformly from the complete candidate pool without replacement. 
    \item \textbf{Stratified Random Sampling.} To control for potential pool-level imbalance, we evaluate a stratified random sample that ensures equal representation across data sources. We sample 1{,}000 pairs from each of the five pools for a total of $k=5{,}000$ pairs. 

\end{itemize}
% \paragraph{Randomized Curation Strategies}

% To isolate the contribution of quality filtering and semantic similarity-based selection, we compare against two randomized sampling baselines that operate directly on the candidate pool without applying any filtering, deduplication, or embedding-based ranking. 

    % This baseline provides a lower bound on performance by selecting data without any quality assessment or semantic guidance. It tests whether the observed improvements are from the curation mechanisms themselves or would occur with any arbitrary subset of data.

    %This baseline removes the confound of pool imbalance while avoiding quality or semantic filtering, allowing us to isolate whether performance differences arise from pool composition versus the quality and relevance of selected examples. 
% \end{itemize}

% By comparing against these baselines, we can quantify the individual contribution of annotation-based quality filtering, semantic ranking, and our proposed SNS and EEE methods to downstream performance. 
\subsection{Downstream Performance Experiments}

Figure~\ref{fig:val_perplexity} and Table~\ref{tab:val_perplexity} present the validation perplexity trajectories for all seven data curation strategies, each averaged over three independent training runs with 95\% confidence intervals. At step~750, the lowest perplexity of $7.732$ is achieved by both \emph{SNS Bidirectional + EEE + Projection} and \emph{SNS Backward + EEE + Projection}. 

A clear three-tier pattern emerges among the methods. First, the \emph{Traditional Data Curation Baseline} is consistently and significantly worse than all other methods, converging to $7.781 \pm 0.003$ at step~750---well above the rest of the field. Second, the \emph{EEE + Projection Only} variant and the two random baselines form a middle tier, converging to perplexities between $7.736$ and $7.752$. Third, the three SNS-augmented variants achieve the lowest perplexities, with \emph{SNS Backward + EEE + Projection} and \emph{SNS Bidirectional + EEE + Projection} reaching $7.734$ and $7.732$ respectively at step~750.

\begin{table}[t]
\centering
\caption{Modality gap ($\ell_2$ distance between per-modality embedding centroids) for each expert encoder and the learned projection network. The projection network reduces the average modality gap by over \textit{90\%} relative to the base experts. Figure \ref{fig:embed-space} depicts the embedding space geometry.}
\label{tab:modality-gaps-main}
\begin{tabular}{@{}lcccc|cccc@{}}
\toprule
 & \multicolumn{4}{c|}{\textbf{Gap}} & \multicolumn{4}{c}{\textbf{$\Delta$ vs.\ Projection (\%)}} \\
\textbf{Space} & \textbf{Video} & \textbf{Audio} & \textbf{Image} & \textbf{Text} & \textbf{Video} & \textbf{Audio} & \textbf{Image} & \textbf{Text} \\
\midrule
E2E Expert      & 44.29 & 44.79 & 27.59 & 18.77 & $-$99.8 & $-$99.8 & $-$99.7 & $-$99.6 \\
Fusion Expert   & 30.72 & 31.21 & 30.80 & 46.13 & $-$99.7 & $-$99.7 & $-$99.7 & $-$99.8 \\
Text Expert     &  0.440 &  0.292 &  0.263 &  0.254 & $-$81.6 & $-$66.6 & $-$66.8 & $-$71.6 \\
\textit{EEE + Projection (Ours)}  &  \textbf{0.081} &  \textbf{0.098} &  \textbf{0.087} &  \textbf{0.072} & --- & --- & --- & --- \\
\bottomrule
\end{tabular}
\end{table}

% Crucially, while the EEE projection alone does not improve over the random baselines, combining it with SNS in any direction yields consistent gains.
Table \ref{tab:modality-gaps-main} illustrates the modality gap collapse effect from our approach (\textit{EEE + Projection}) compared to base experts, reducing $\ell_2$ distance between per-modality centroids by over 90\% on average.

\section{Conclusion}

Ranked retrieval is increasingly used to curate training mixtures across multimodal, multidomain datasets. However, our study shows that small geometric quirks and label noise can compound into systematic errors when retrieval is treated as standard infrastructure. 
%A traditional baseline, which applies heuristic text‑only filtering and single‑encoder annotation ranking, yields a validation perplexity of 7.780, significantly worse than stratified random sampling (7.737).
We demonstrate that standard retrieval based data selection can actively degrade training mixtures when the embedding space exhibits uncorrected modality bias. 

We presented a framework that addresses this by acting jointly on training pairs and embeddings through two complementary components: Symmetric Nucleus Subsampling (SNS) and Expert Embedding Engine (EEE). The SNS reduces raw data to supervision misalignment by trimming the raw input and the annotation under a symmetric similarity gate. The EEE trains a mixture of embedding experts with a learned projection using a bias-aware objective. 

In our multimodal instruction-following evaluation on a five-pool mixture spanning audio-text, image-text, video-text, and text-text pairs, the \emph{SNS Bidirectional + EEE} configuration achieved the lowest validation perplexity score of \textbf{7.732} outperforming stratified random sampling and the traditional data curation pipeline. 
%Ablations show that EEE alone roughly matches random baselines, while adding the SNS in any direction yields consistent gains, with the bidirectional SNS demonstrating the greatest improvement. These findings support the view that pair-level denoising and geometry-level debiasing are complementary. 
This performance is underpinned by a fundamental geometric shift: our projection network collapses the modality gap, reducing the $\ell_2$ separation between modality centroids by an average of over \textbf{90\%} compared to base experts (Table \ref{tab:modality-gaps-main}), ensuring that retrieval is driven by semantic alignment rather than input format.

Despite these gains, our framework has limitations. First, our multimodal instruction-following evaluation is limited in scale.
%as we use 500 human-curated samples focused on a single query relating to natural scenes. It is possible that the observed gains may not fully capture behavior in other domains or at scale.
Secondly, our traditional curation baseline is limited as heuristic quality filters and single encoder rankings are applied to text-based supervisions only, which may underestimate what a stronger multimodal curation pipeline operating on both the raw data and supervision could achieve. 

%next steps%
A natural next step is to make both trimming and projection uncertainty-aware, using signals like expert disagreement, augmentation instability, or description inconsistency to decide when to trim, when to fall back, and which experts to trust. Another direction is to decouple proposal from verification: use fast embeddings to retrieve candidates, then apply a stronger cross-modal verifier for reranking to reduce self-reinforcing selection effects. Finally, we plan to evaluate curation more directly and at larger scales, with metrics that track mixture diversity and tail coverage over iterations.

%, and with auditing that monitors privacy, licensing, and distribution shifts introduced by retrieval-based selection.

%closing% 
\clearpage
\subsubsection*{Acknowledgments}

\paragraph{LLM Usage (manuscript wide).}In accordance with ICLR 2026 policy, we disclose that we used a large language model during manuscript preparation across multiple sections of the paper. The model was used to assist with drafting and editing text, improving clarity and organization, and suggesting alternative phrasing. All technical content, claims, experimental results, and citations were reviewed and verified by the authors, who take full responsibility for the final manuscript.

\bibliography{iclr2026_conference}
\bibliographystyle{iclr2026_conference}

\clearpage

\appendix
\section{Appendix}

\subsection{SNS ablations}
\label{ref:sns_ablations}

In the following, we ablate the Symmetric Nucleus Subsampler (SNS) component across various hyperparameter ranges and configurations.

\subsubsection{SNS definition}
\paragraph{Forward extraction.}
The forward extraction function $\mathcal{N}_{\alpha}$ identifies raw components most relevant to the annotation:
\begin{equation}
\tilde{x} = \mathcal{N}_{\alpha}(x, y) = \{x_i : I(x_i; y) > \tau_{\alpha}\},
\label{eq:sns_forward}
\end{equation}
where $x_i$ denotes components such as text spans, image regions, or video clips, and $\tau_{\alpha}$ controls extraction granularity.

\paragraph{Backward extraction.}
Symmetrically, $\mathcal{N}_{b}$ retains only annotation components grounded in observable data:
\begin{equation}
\tilde{y} = \mathcal{N}_{b}(x, y) = \{y_j : I(x; y_j) > \tau_{b}\},
\label{eq:sns_backward}
\end{equation}
which removes label or tag elements that are not supported by the raw input.

\paragraph{Information density.}
The nucleus pair $(\tilde{x}, \tilde{y})$ increases information density:
\begin{equation}
\rho(\tilde{x}, \tilde{y}) = \frac{I(\tilde{x}; \tilde{y})}{|\tilde{x}| + |\tilde{y}|},
\label{eq:info_density}
\end{equation}
since extraction aims to preserve $I(\tilde{x}; \tilde{y}) \approx I(x; y)$ while reducing $|\tilde{x}| + |\tilde{y}|$.

\subsubsection{SNS hyperparameters}
We ablate three SNS knobs.

\begin{itemize}
    \item \textbf{Directionality.} Forward (Eq.~\ref{eq:sns_forward}), backward (Eq.~\ref{eq:sns_backward}), or bidirectional extraction. With reinjection enabled, SNS can overwrite either $x$ or $y$ with $(\tilde{x}, \tilde{y})$ when the gate accepts the variant. We enable reinjection mode for all experiments.
    \item \textbf{MI gate ratio $\rho$.} We compute $\mathrm{sim}(\tilde{x}, \tilde{y})$ using a unified multimodal encoder and accept variants if
    \begin{equation}
    \mathrm{sim}(\tilde{x}, \tilde{y}) \ge \rho \cdot \mathrm{sim}(x, y).
    \label{eq:mi_gate}
    \end{equation}
    Larger $\rho$ is stricter and requires variants to match or exceed the original alignment.
    \item \textbf{Thresholds $\tau_{\alpha}$ and $\tau_{b}$.} These control extraction granularity in the forward and backward directions. Lower thresholds yield larger nuclei, while higher thresholds produce more focused nuclei.
\end{itemize}

\clearpage

\subsubsection{Direction ablation}
Below are results from varying SNS directionality (\texttt{OFF}, \texttt{FORWARD}, \texttt{BACKWARD}, \texttt{BIDIRECTIONAL}) with $\rho = 1.00$ for 350 paired samples for each of the 5 data pools.

\begin{figure}[H]
\centering
\includegraphics[width=0.75\linewidth]{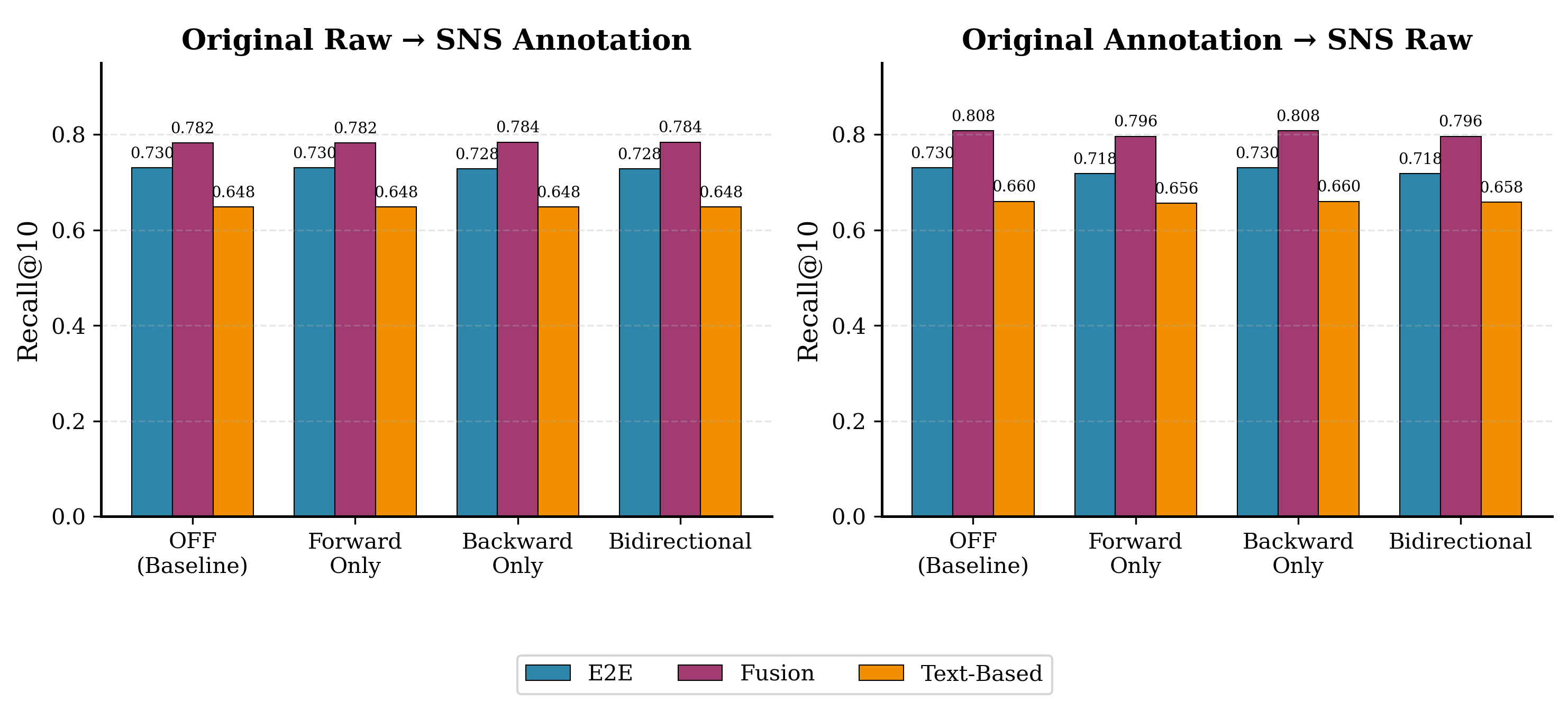}
\vspace{-0.5em}
\caption{R@10 versus SNS directionality for both retrieval directions: Annotation to Raw Data (A2R); Raw Data to Annotation (R2A).}
\label{fig:direction}
\end{figure}

\begin{table}[H]
\centering
\caption{Direction ablation of A2R \& R2A R@10 by SNS configuration ($\rho=1.00$).}
\label{tab:direction_ablation}
\begin{tabular}{@{}l|ccc|ccc@{}}
\toprule
& \multicolumn{3}{c|}{\textbf{R$\rightarrow$A}} & \multicolumn{3}{c}{\textbf{A$\rightarrow$R}} \\
\textbf{Configuration} & E2E & Fusion & Text & E2E & Fusion & Text \\
\midrule
Baseline (OFF) & \textbf{0.7300} & 0.7820 & 0.6480 & \textbf{0.7300} & 0.8080 & \textbf{0.6600} \\
Forward Only   & \textbf{0.7300} & 0.7820 & 0.6480 & 0.7180 & 0.7960 & 0.6560 \\
Backward Only & 0.7280 & \textbf{0.7840} & 0.6480 & \textbf{0.7300} & \textbf{0.8080} & \textbf{0.6600} \\
Bidirectional  & 0.7280 & 0.7840 & 0.6480 & 0.7180 & 0.7960 & 0.6580 \\
\bottomrule
\end{tabular}
\end{table}

\begin{figure}[H]
\centering
\includegraphics[width=0.85\linewidth]{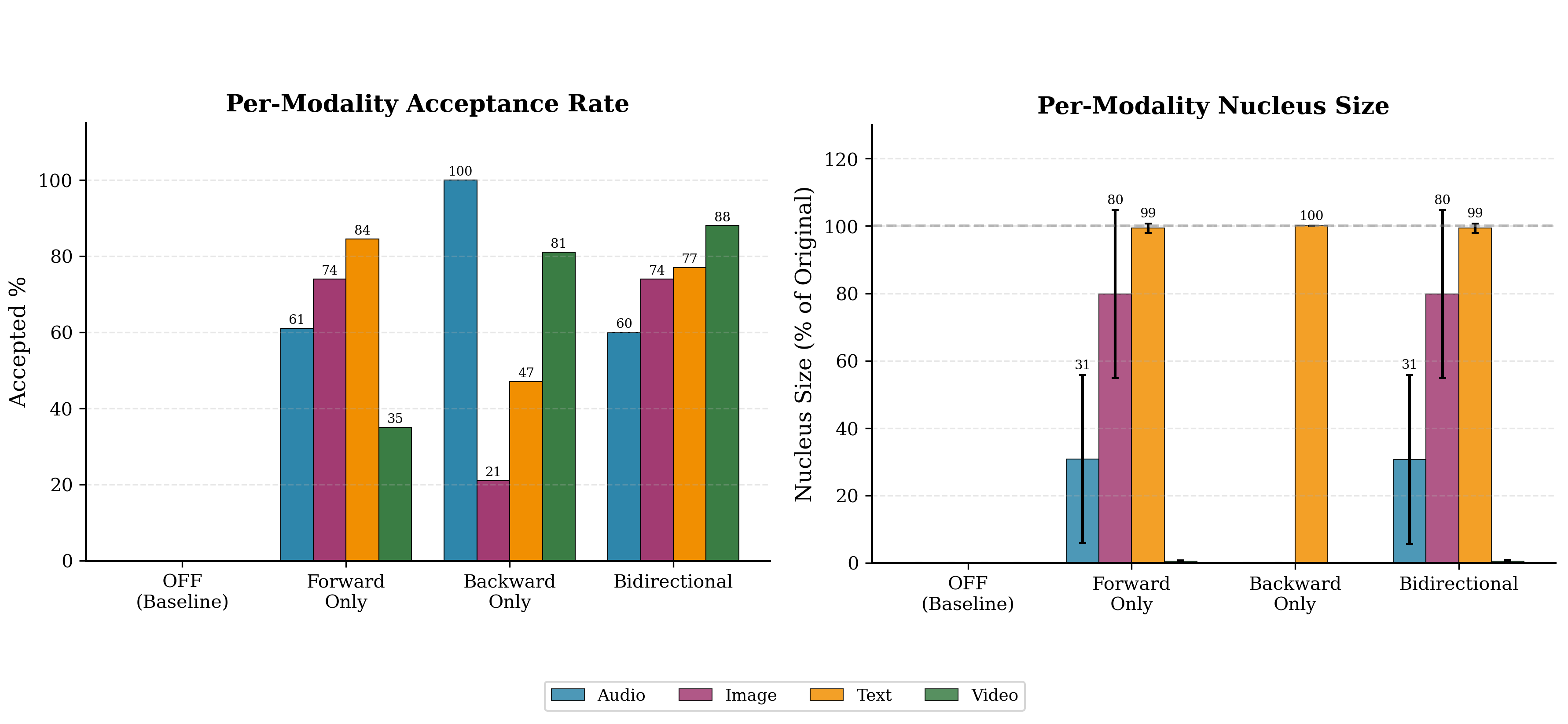}
\vspace{-0.5em}
\caption{\textit{Left}: fraction of accepted nucleus samples (fixed $\rho=1.00$). \textit{Right}: accepted nucleus size difference in bytes (\textit{note}: for backwards extraction - all annotations are text).}
\label{fig:direction_frac}
\end{figure}

\clearpage

\subsubsection{MI gate ablation}
Below are results from varying the MI gate ratio $\rho$ on a fixed sample of 1{,}000 pairs (200 per pool).

\begin{figure}[H]
\centering
\includegraphics[width=0.85\linewidth]{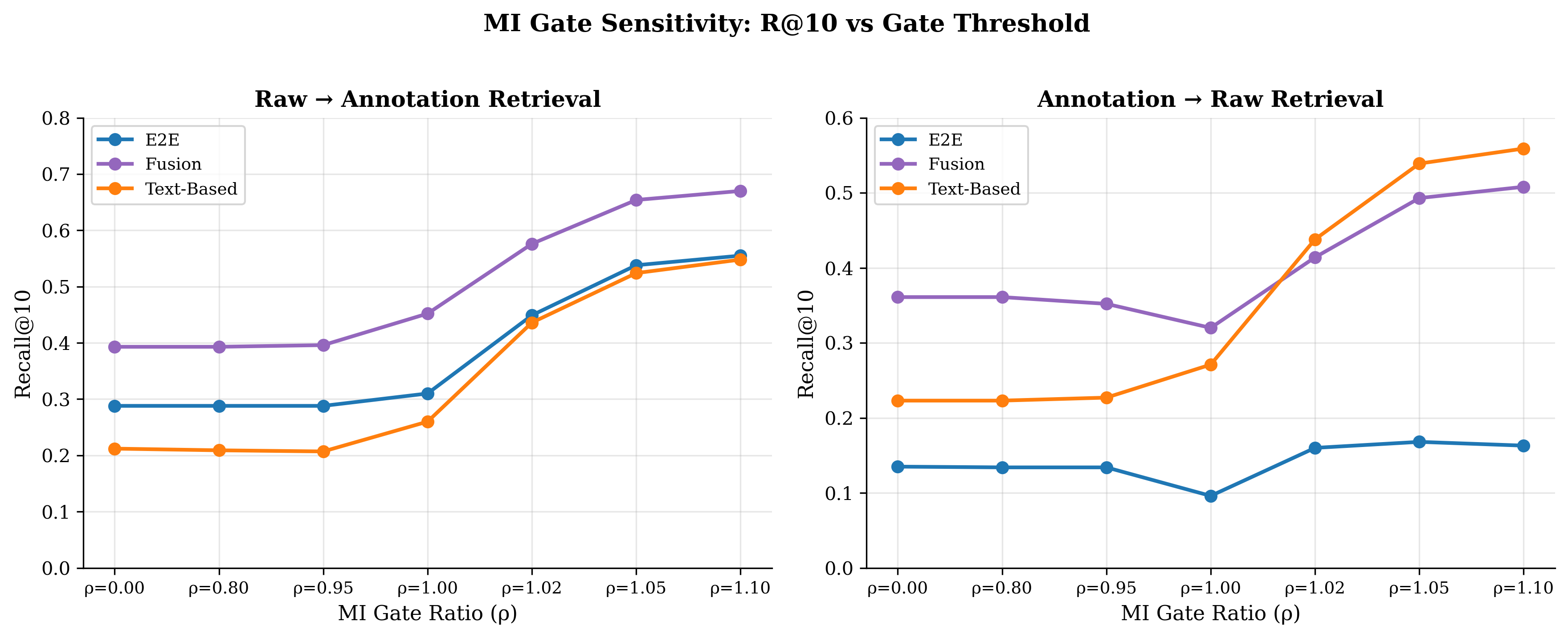}
\vspace{-0.5em}
\caption{A2R \& R2A R@10 versus MI gate ratio $\rho$.}
\label{fig:mi}
\end{figure}

\begin{figure}[H]
\centering
\includegraphics[width=0.75\linewidth]{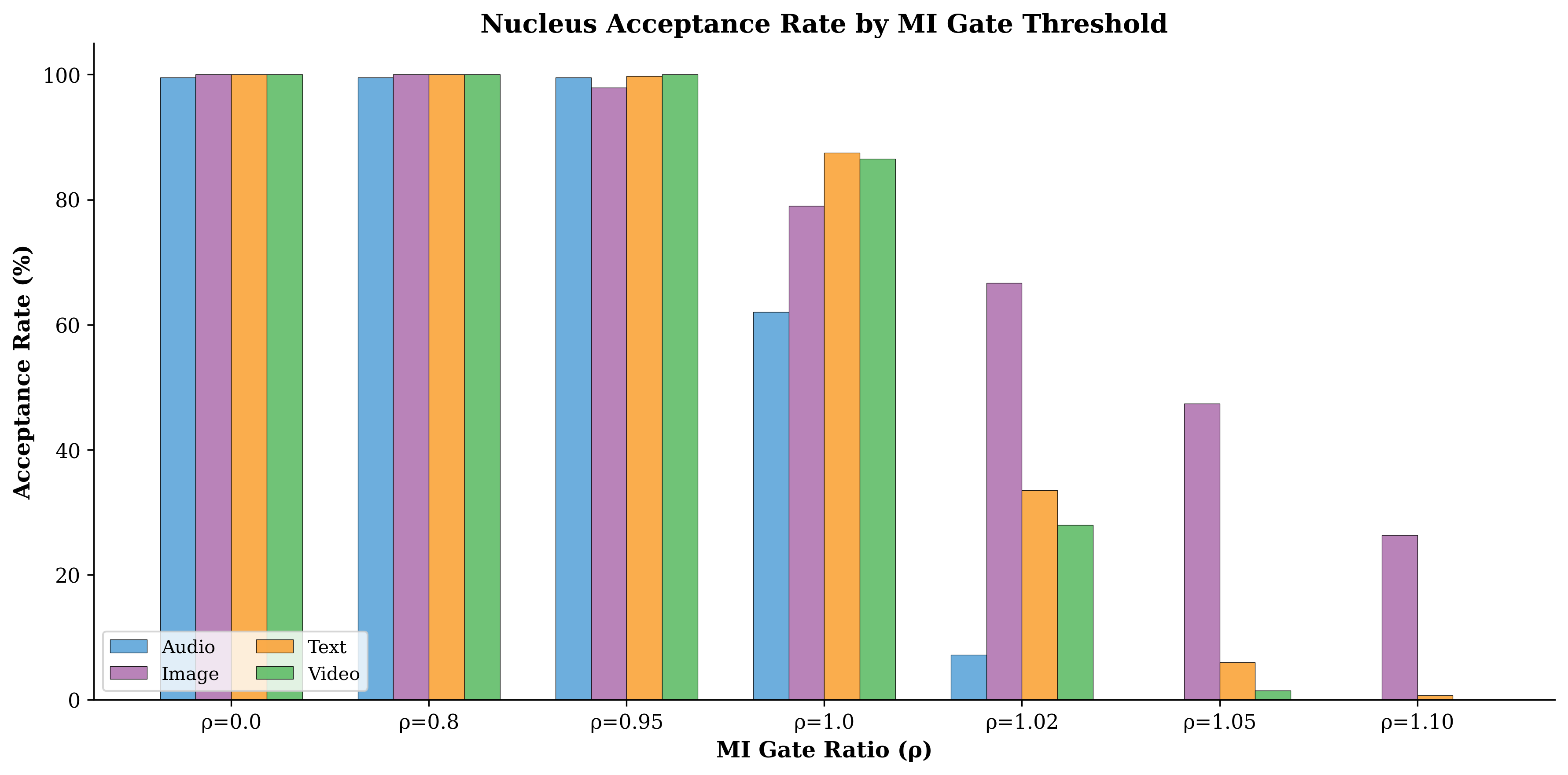}
\vspace{-0.5em}
\caption{Acceptance rate versus MI gate ratio $\rho$.}
\label{fig:accept}
\end{figure}

At higher $\rho$, retrieval improves while the acceptance rate decreases (Figure~\ref{fig:mi} and Figure~\ref{fig:accept}). This indicates a precision-oriented effect where stricter gating yields fewer but higher alignment pairs. In addition to R@10, a complementary analysis of nucleus compactness and cross-modality neighborhood consistency can better reflect the information density objective of SNS.

\begin{table}[H]
\centering
\caption{MI gate ablation: R@10 by expert and retrieval direction.}
\label{tab:mi}
\textbf{(a) Raw $\rightarrow$ Annotation Retrieval}\vspace{0.3em}

\begin{tabular}{@{}lccccccc@{}}
\toprule
\textbf{Expert} & $\rho$=0.0 & $\rho$=0.80 & $\rho$=0.95 & $\rho$=1.00 & $\rho$=1.02 & $\rho$=1.05 & $\rho$=1.10 \\
\midrule
End-to-End & 0.29 & 0.29 & 0.29 & 0.31 & 0.45 & 0.54 & \textbf{0.56} \\
Fusion & 0.39 & 0.39 & 0.40 & 0.45 & 0.58 & 0.65 & \textbf{0.67} \\
Text-Based & 0.21 & 0.21 & 0.21 & 0.26 & 0.44 & 0.52 & \textbf{0.55} \\
\bottomrule
\end{tabular}

\vspace{1em}

\textbf{(b) Annotation $\rightarrow$ Raw Retrieval}\vspace{0.3em}

\begin{tabular}{@{}lccccccc@{}}
\toprule
\textbf{Expert} & $\rho$=0.0 & $\rho$=0.80 & $\rho$=0.95 & $\rho$=1.00 & $\rho$=1.02 & $\rho$=1.05 & $\rho$=1.10 \\
\midrule
End-to-End & 0.14 & 0.13 & 0.13 & 0.10 & 0.16 & \textbf{0.17} & 0.16 \\
Fusion & 0.36 & 0.36 & 0.35 & 0.32 & 0.41 & 0.49 & \textbf{0.51} \\
Text-Based & 0.22 & 0.22 & 0.23 & 0.27 & 0.44 & 0.54 & \textbf{0.56} \\
\bottomrule
\end{tabular}
\end{table}

\subsubsection{Tau threshold ablation}

As referenced in the main article, we ablate SNS with modality-specific $\tau_\alpha$ and $\tau_\beta$. In practice, we have modality-specific $\tau_\alpha$ and $\tau_\beta$, as we learned through the below experiment that multimodal datasets have modality-specific embedding similarity (a proxy to mutual information density) ranges. In the following, we compute the pairwise embedding similarity scores for 1024 pairs for each the 4 distinct modality data pools in our dataset.

\begin{figure}[H]
    \centering
    \includegraphics[width=0.5\linewidth]{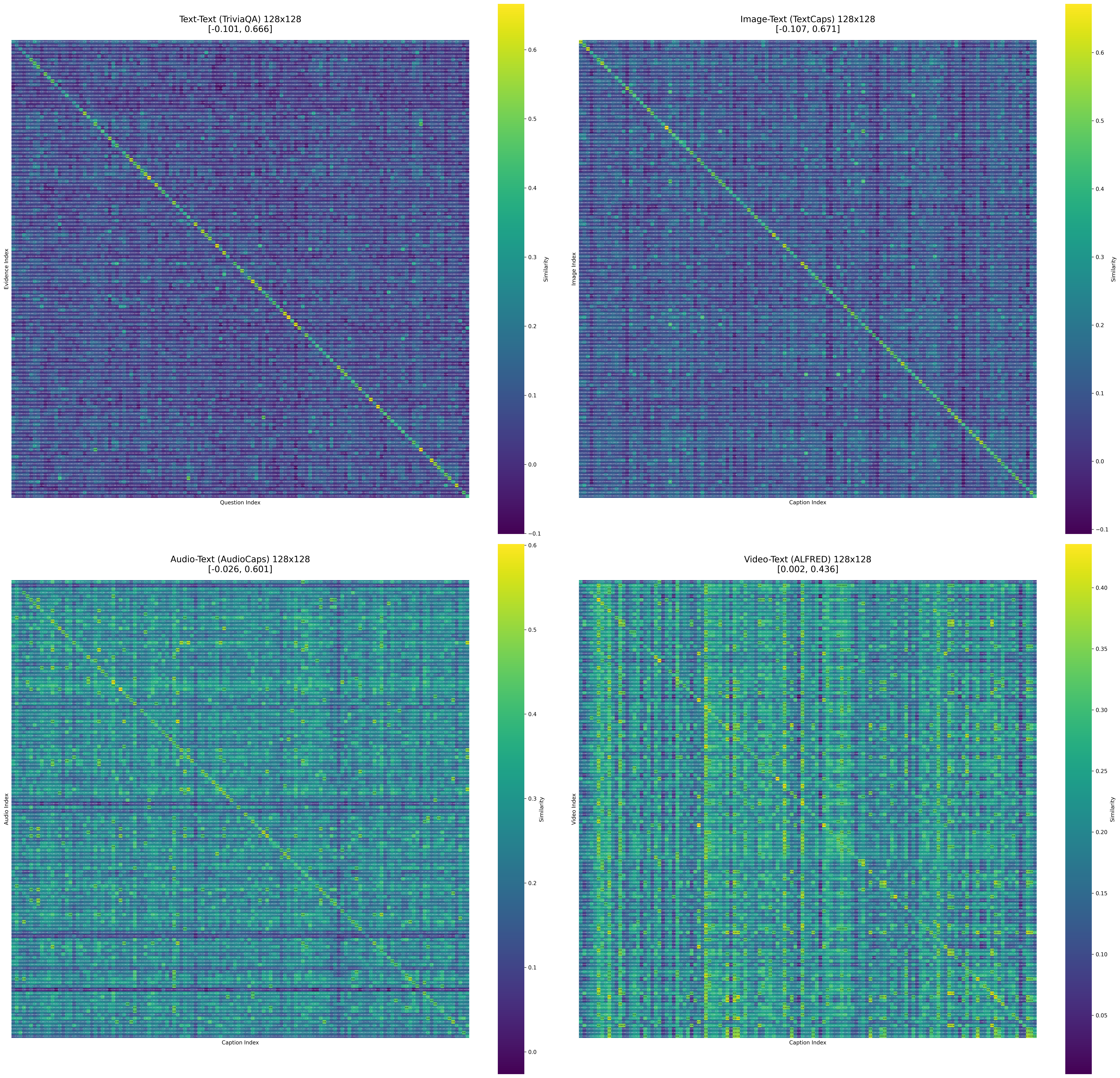}
    \caption{Pairwise similarity across all modalities of paired samples, evaluated on 1024 samples using the Omni-Embed-Nemotron-3B model.}
    \label{fig:pairwise_sim}
\end{figure}

Next, forward thresholds $\tau_{\alpha}$ and backward thresholds $\tau_{b}$ are jointly varied across 250 samples per data pool for the five data pools.

\begin{figure}[H]
\centering
\includegraphics[width=0.95\linewidth]{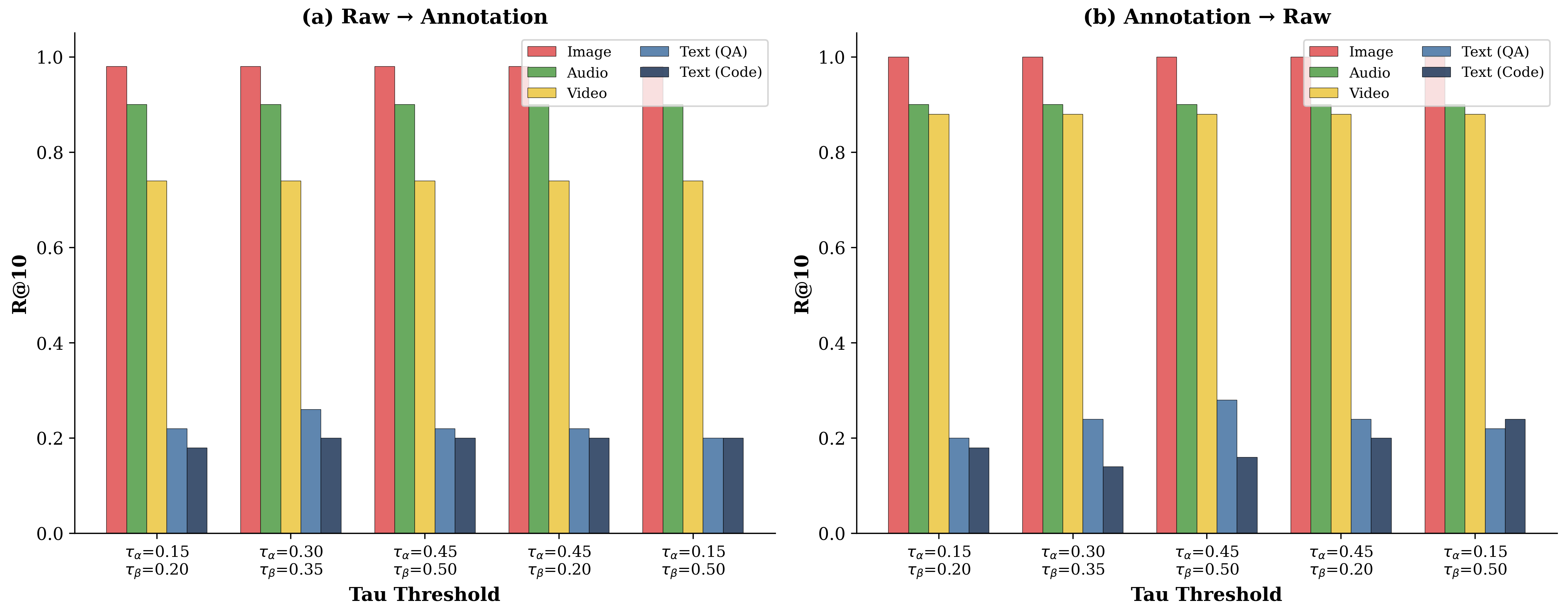}
\vspace{-0.5em}
\caption{A2R \& R2A R@10 versus $(\tau_{\alpha}, \tau_{b})$, stratified by modality.}
\label{fig:tau}
\end{figure}

Jointly varying $\tau_{\alpha}$ and $\tau_{b}$ has minimal impact on retrieval in this range. This suggests components near the decision boundary contribute less to the gated similarity objective, and future work can explore more aggressive thresholds or adaptive thresholding by modality.

\clearpage

\subsection{EEE ablations}

Below, we discuss various ablation experiments applied to the Expert Embedding Engine component.

\subsubsection{Metrics}
We measure cross-modal retrieval using Recall@K (R@K), defined as the fraction of queries where the correct match appears in the top K results:
\begin{equation}
\mathrm{R@K} = \frac{1}{N} \sum_{i=1}^{N} \mathbf{1}\big[\mathrm{rank}(x_i, y_i) \le K\big],
\label{eq:rk}
\end{equation}
where $\mathrm{rank}(x_i, y_i)$ is the position of the true match when retrieving $y_i$ from query $x_i$. We evaluate both directions:
\begin{itemize}
    \item \textbf{R2A (Raw $\to$ Annotation)}: Given raw data, retrieve the matching text annotation.
    \item \textbf{A2R (Annotation $\to$ Raw)}: Given text annotation, retrieve the matching raw data.
\end{itemize}

\subsubsection{Modality gap diagnostic}
Multimodal embeddings can cluster by modality rather than semantic content. For embedding function $f : \bigcup_{m \in \mathcal{M}} \mathcal{X}_m \to \mathbb{R}^d$, modality clustering occurs when:
\begin{equation}
\mathbb{E}\big[\|f(x_i) - f(x_j)\| \mid x_i, x_j \in \mathcal{X}_m\big] <
\mathbb{E}\big[\|f(x_i) - f(x_k)\| \mid x_i \in \mathcal{X}_{m_1}, x_k \in \mathcal{X}_{m_2}\big].
\label{eq:modality_gap}
\end{equation}

\subsubsection{Modality and SNS variant ablations}
We evaluate three expert embedding engines (\textbf{E2E}, \textbf{Fusion}, \textbf{Text}) across five pools with four SNS variants (Baseline, Forward, Backward, Bidirectional).

\begin{figure}[H]
    \centering
    \includegraphics[width=\linewidth]{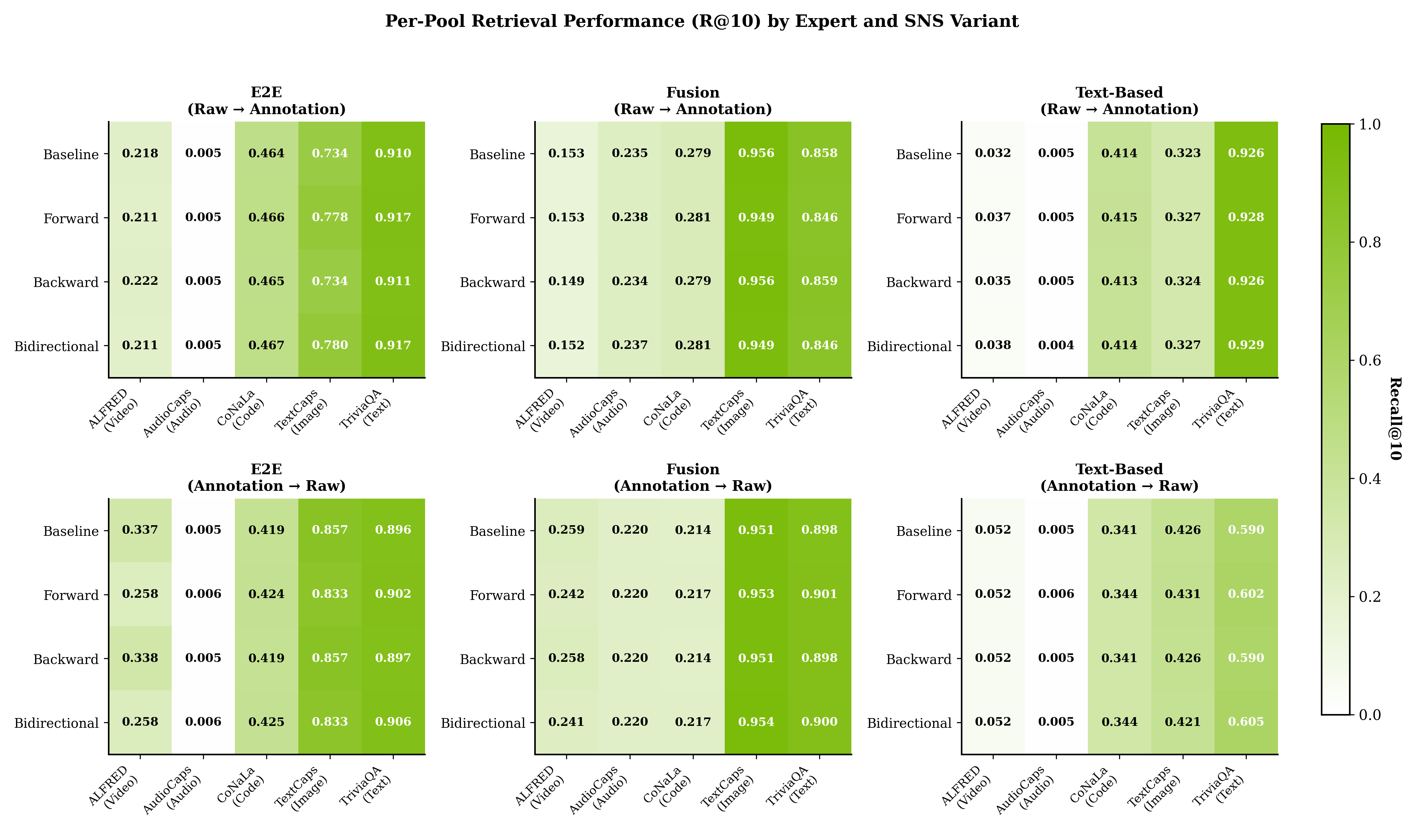}
    \caption{Per pool R@10 heatmaps for A2R and R2A retrieval.}
    \label{fig:pool_heatmaps}
\end{figure}

\begin{figure}[H]
    \centering
    \includegraphics[width=0.75\linewidth]{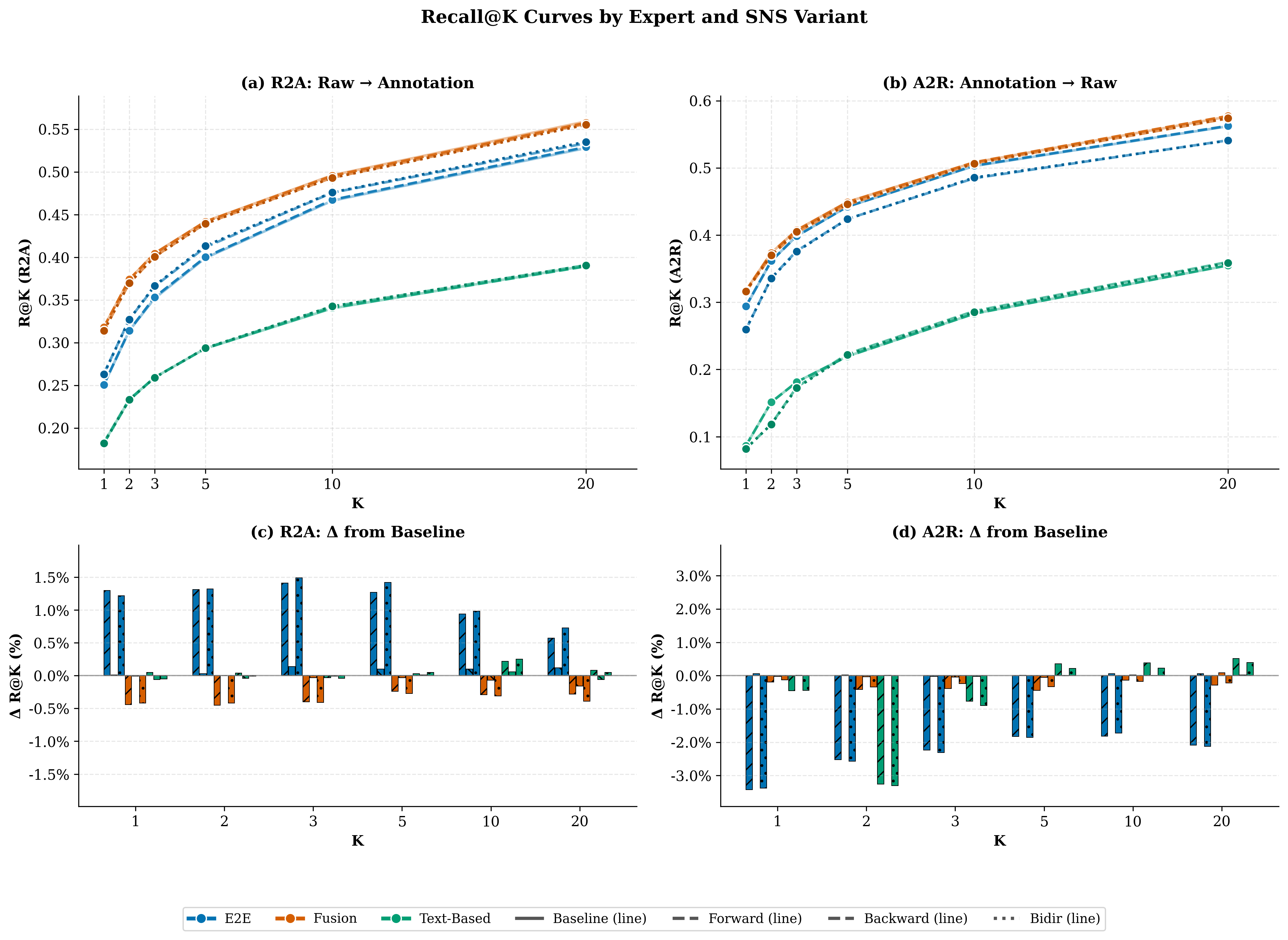}
    \caption{Recall@K curves and change from baseline for each expert and SNS variant.}
    \label{fig:recall_curves}
\end{figure}

\begin{table}[H]
\centering
\caption{Recall@K by expert and SNS variant. Values shown as R2A\,/\,A2R.}
\label{tab:recall_k}
\footnotesize
\setlength{\tabcolsep}{4pt}
\renewcommand{\arraystretch}{0.9}
\begin{tabular}{@{}llcccccc@{}}
\toprule
\textbf{Expert} & \textbf{SNS} & \textbf{R@1} & \textbf{R@2} & \textbf{R@3} & \textbf{R@5} & \textbf{R@10} & \textbf{R@20} \\
\midrule
\multirow{4}{*}{\textbf{E2E}}
& Baseline & .251/.294 & .314/.361 & .352/.399 & .399/.442 & .466/.503 & .528/.562 \\
& Forward  & .264/.259 & .327/.336 & .366/.376 & .412/.424 & .476/.485 & .534/.541 \\
& Backward & .251/.294 & .314/.362 & .353/.399 & .400/.442 & .467/.503 & .529/.563 \\
& Bidir    & .263/.260 & .327/.336 & .367/.376 & .413/.424 & .476/.486 & .535/.541 \\
\midrule
\multirow{4}{*}{\textbf{Fusion}}
& Baseline & \textbf{.318/.318} & \textbf{.374/.374} & \textbf{.405/.408} & \textbf{.442/.450} & .496/.508 & .559/.576 \\
& Forward  & .314/.316 & .370/.370 & .401/.404 & .440/.445 & .493/.507 & .556/.574 \\
& Backward & \textbf{.318}/.317 & \textbf{.374}/.373 & .404/.407 & \textbf{.442}/.449 & .495/\textbf{.508} & .558/\textbf{.577} \\
& Bidir    & .314/.316 & .370/.370 & .401/.405 & .439/.446 & .493/.507 & .555/.574 \\
\midrule
\multirow{4}{*}{\textbf{Text}}
& Baseline & .183/.087 & .233/.151 & .259/.181 & .293/.220 & .340/.283 & .390/.355 \\
& Forward  & .183/.082 & .234/.119 & .259/.174 & .293/.223 & .342/.287 & .391/.360 \\
& Backward & .182/.087 & .233/.151 & .259/.181 & .293/.220 & .341/.283 & .389/.355 \\
& Bidir    & .182/.082 & .233/.118 & .259/.172 & .294/.222 & .343/.285 & .391/.359 \\
\bottomrule
\end{tabular}
\end{table}

\begin{figure}[H]
    \centering
    \includegraphics[width=0.75\linewidth]{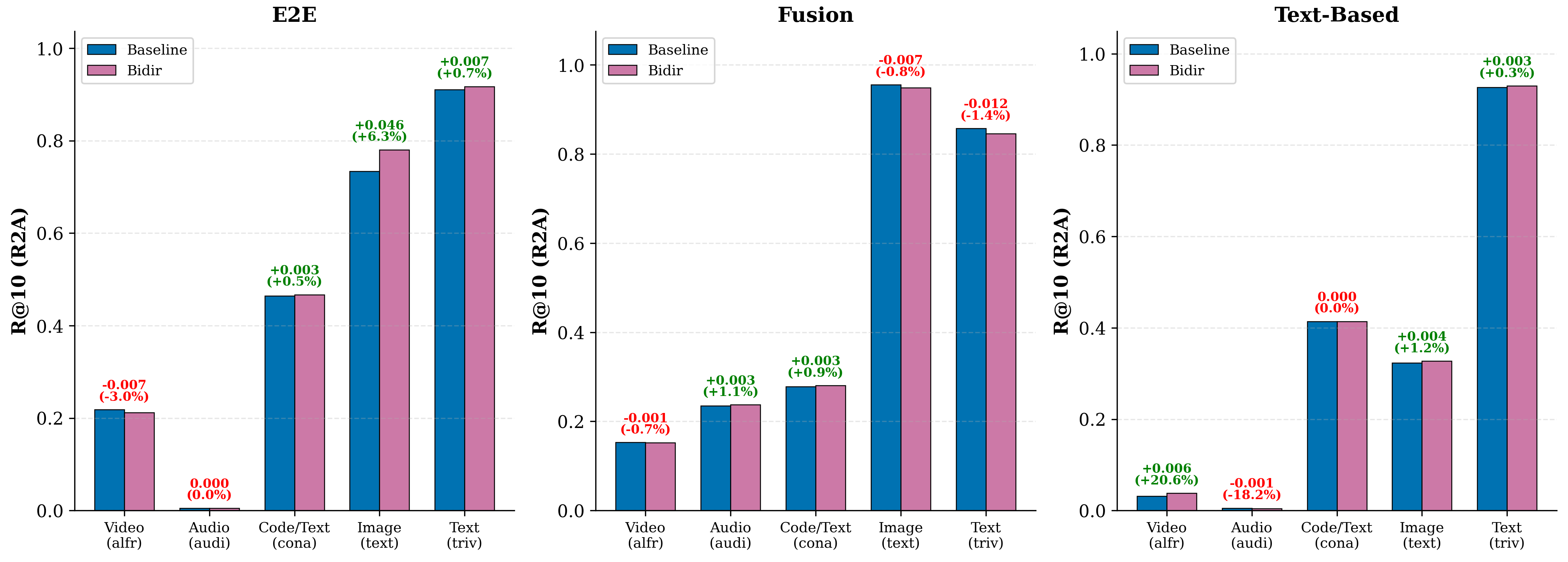}
    \caption{Baseline versus Bidirectional SNS by pool, R2A R@10 retrieval impact}
    \label{fig:baseline_vs_bidir}
\end{figure}

We find that the fusion expert performs well across various R@K values and demonstrates the strongest single-expert correlation between raw data and paired annotation compared to text-based and end-to-end multimodal embedding model implementations.

\clearpage

\subsubsection{Expert Ablations - Downstream Eval Performance}

In the following, we share the isolated downstream evaluation results of the individual experts in the EEE. To isolate the experts, we disable SNS \& the projection network, and evaluate the individual experts' embedding spaces as datablend curation semantic maps. We find that individual experts curate datablends differently, with vastly different modality compositions, likely due to the geometry of the embeddings in the respective expert's embedding spaces.

\begin{figure}[H]
\centering

% Top row: two subfigures
\begin{subfigure}[t]{0.42\textwidth}
\centering
\includegraphics[width=\textwidth]{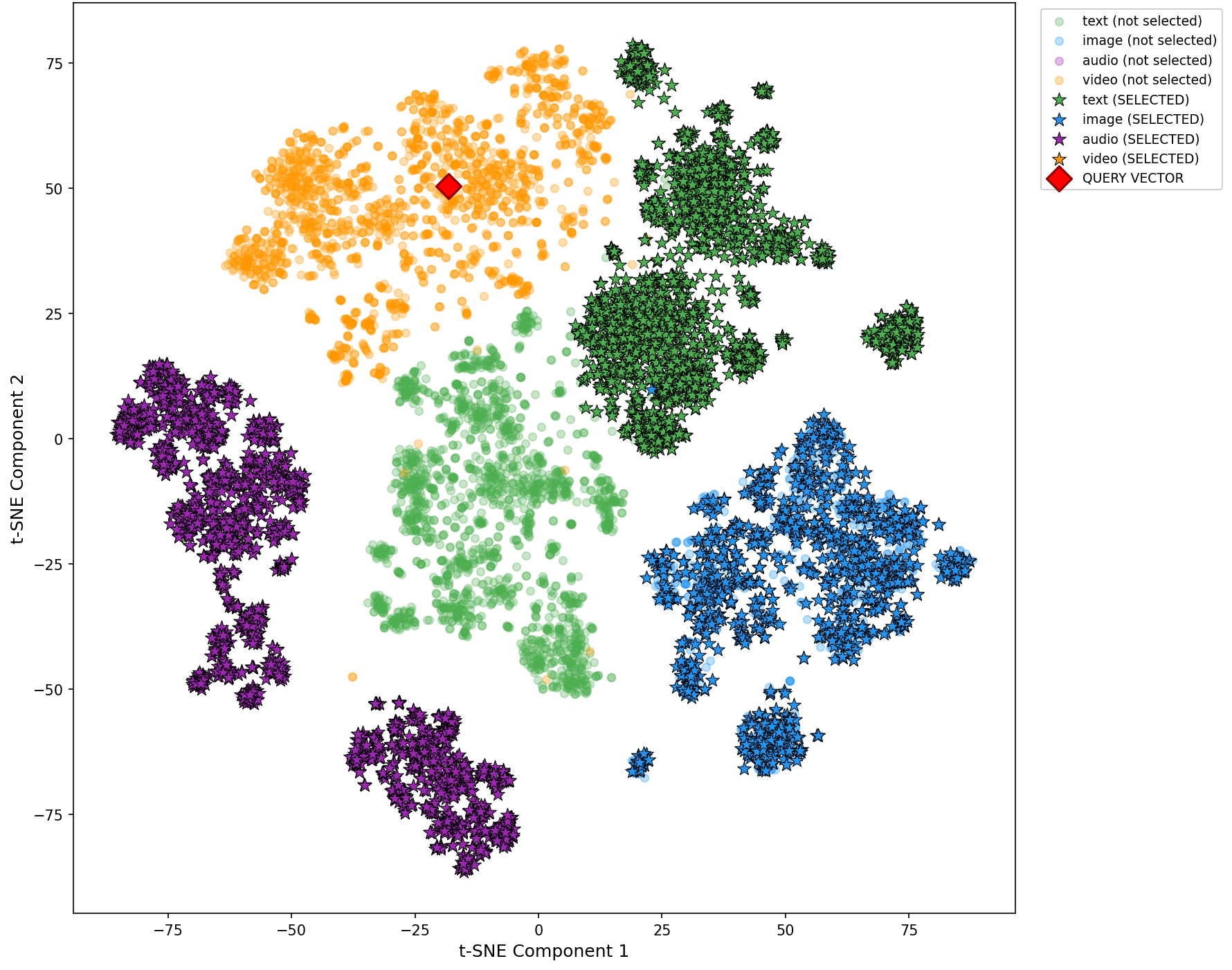}
\caption{\textbf{Text-based expert}}
\label{fig:1}
\end{subfigure}
\hfill
\begin{subfigure}[t]{0.42\textwidth}
\centering
\includegraphics[width=\textwidth]{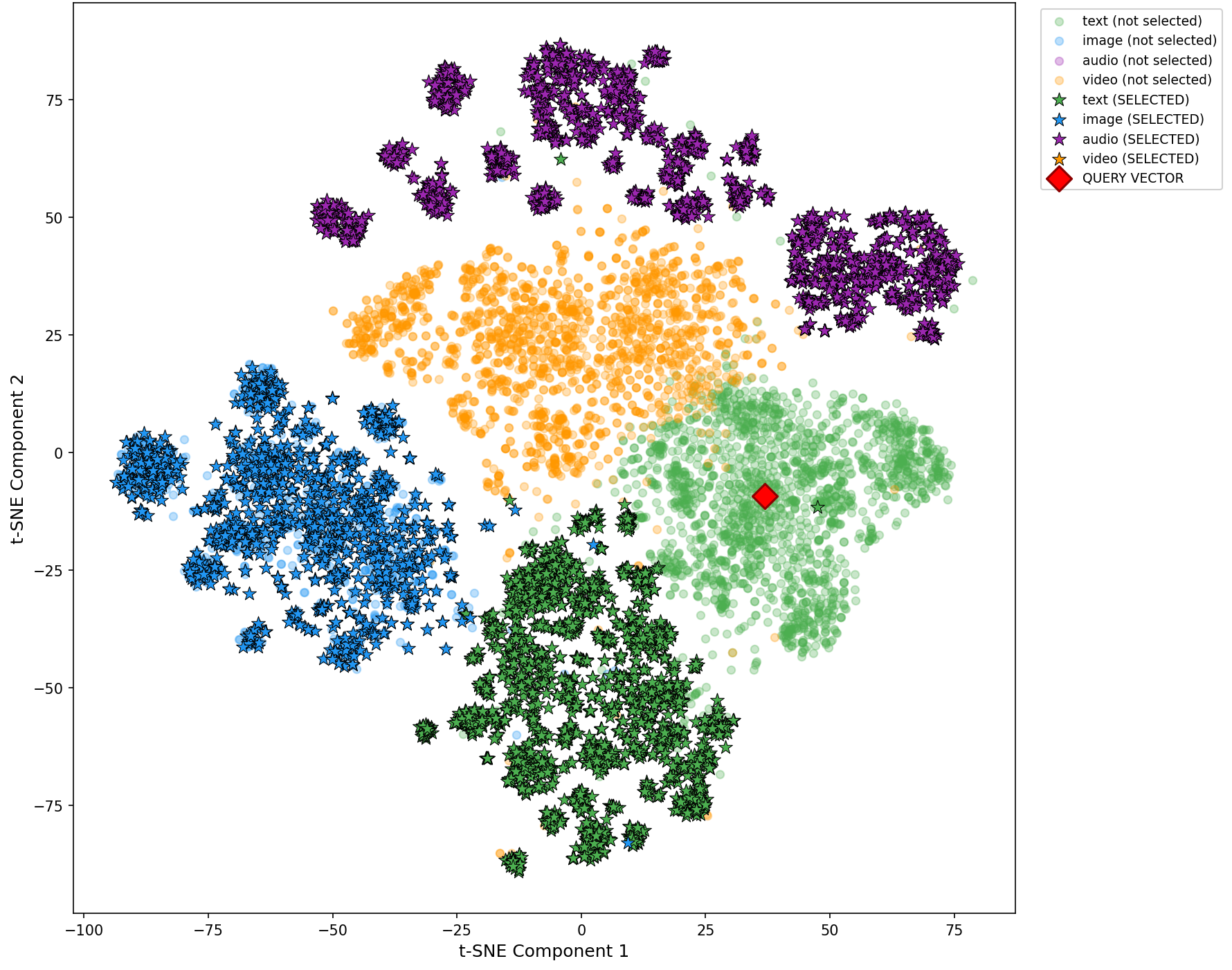}
\caption{\textbf{Fusion expert}}
\label{fig:2}
\end{subfigure}

\vspace{0.5em}  % Minimal vertical spacing between rows

% Bottom row: one subfigure
\begin{subfigure}[t]{\textwidth}
\centering
\includegraphics[width=0.42\textwidth]{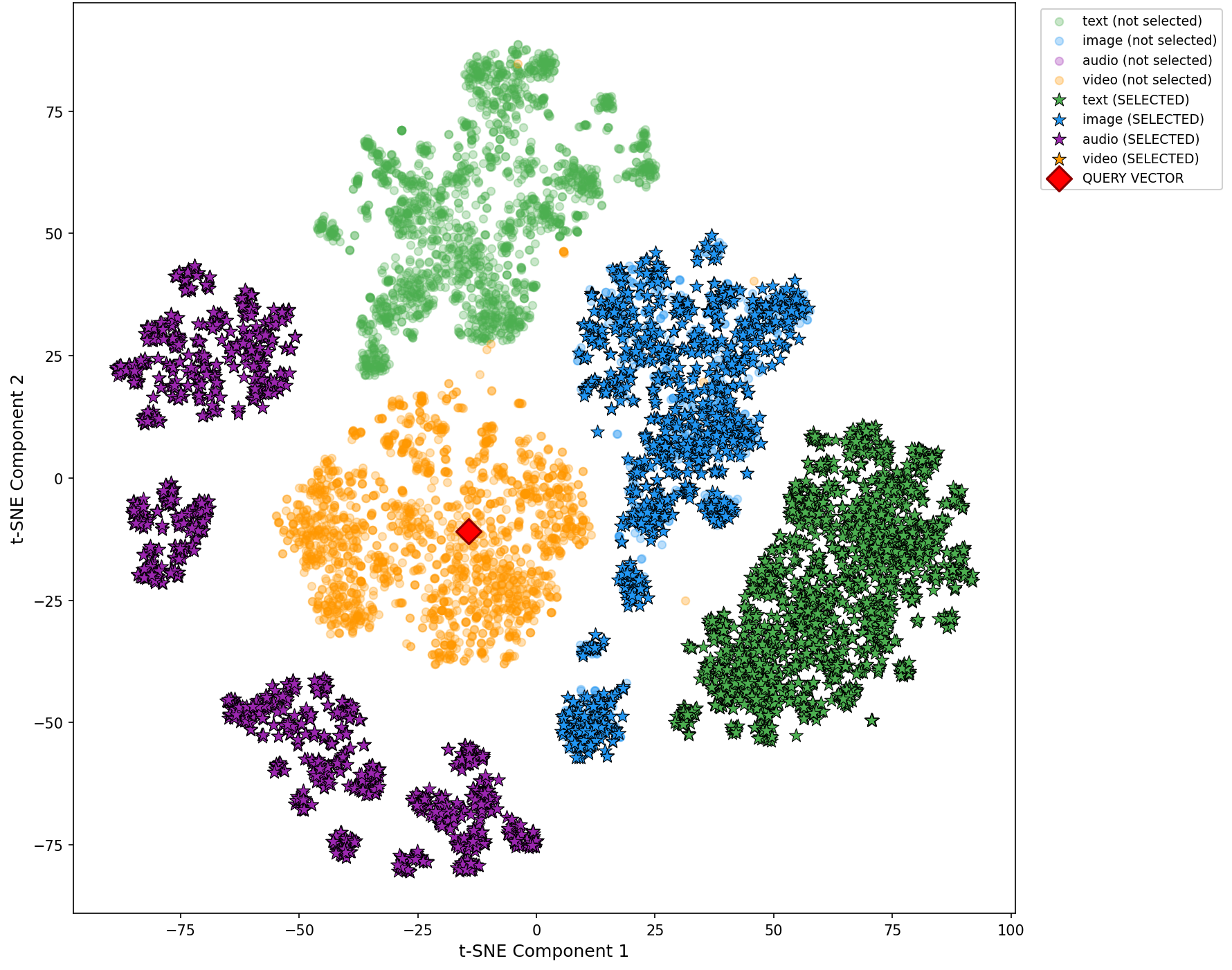}  % Adjust width as needed for centering
\caption{\textbf{End-to-End (E2E) expert}}
\label{fig:3}
\end{subfigure}

\label{fig:overall}
\caption{2D t-SNE visualization of curated samples (5000 samples of 10000) given query vector ``natural, real-world scenes with objects, landscape, subjects, or people'' across isolated EEE experts.}
\end{figure}

\begin{figure}[H]
    \centering
    \includegraphics[width=0.85\linewidth]{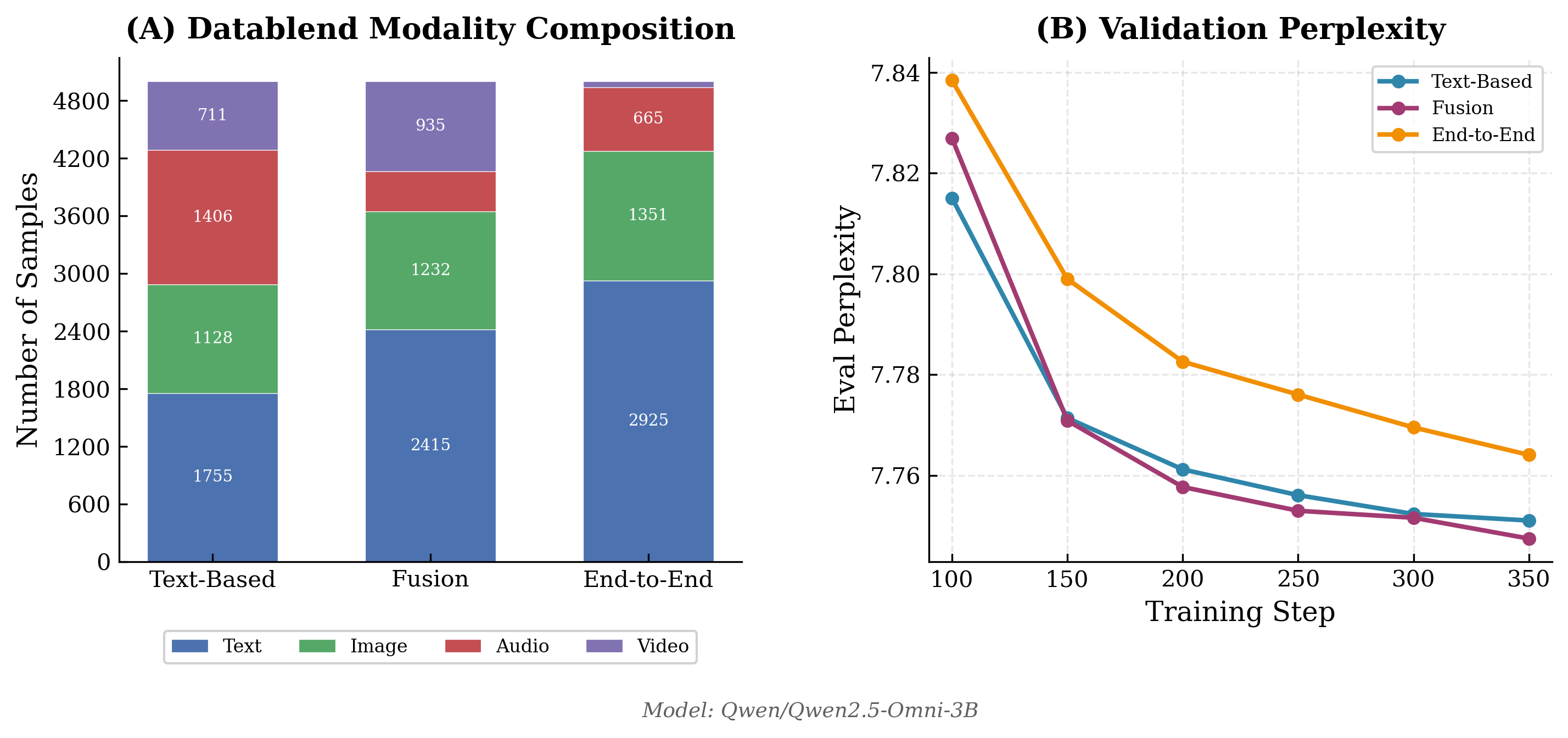}
    \caption{\textit{Left}: Datablend Modality Composition by EEE expert curation, \textit{Right}: Downstream eval validation perplexity curve across 1 epoch fine-tuning Qwen-2.5-Omni-3B for multimodal understanding on curated datablends by EEE expert.}
    \label{fig:datablend-modality-eee}
\end{figure}

\clearpage

\subsection{Projection Network Ablations}
\label{ref:proj-ablations}

\begin{figure}[H]
    \centering
    \includegraphics[width=0.65\linewidth]{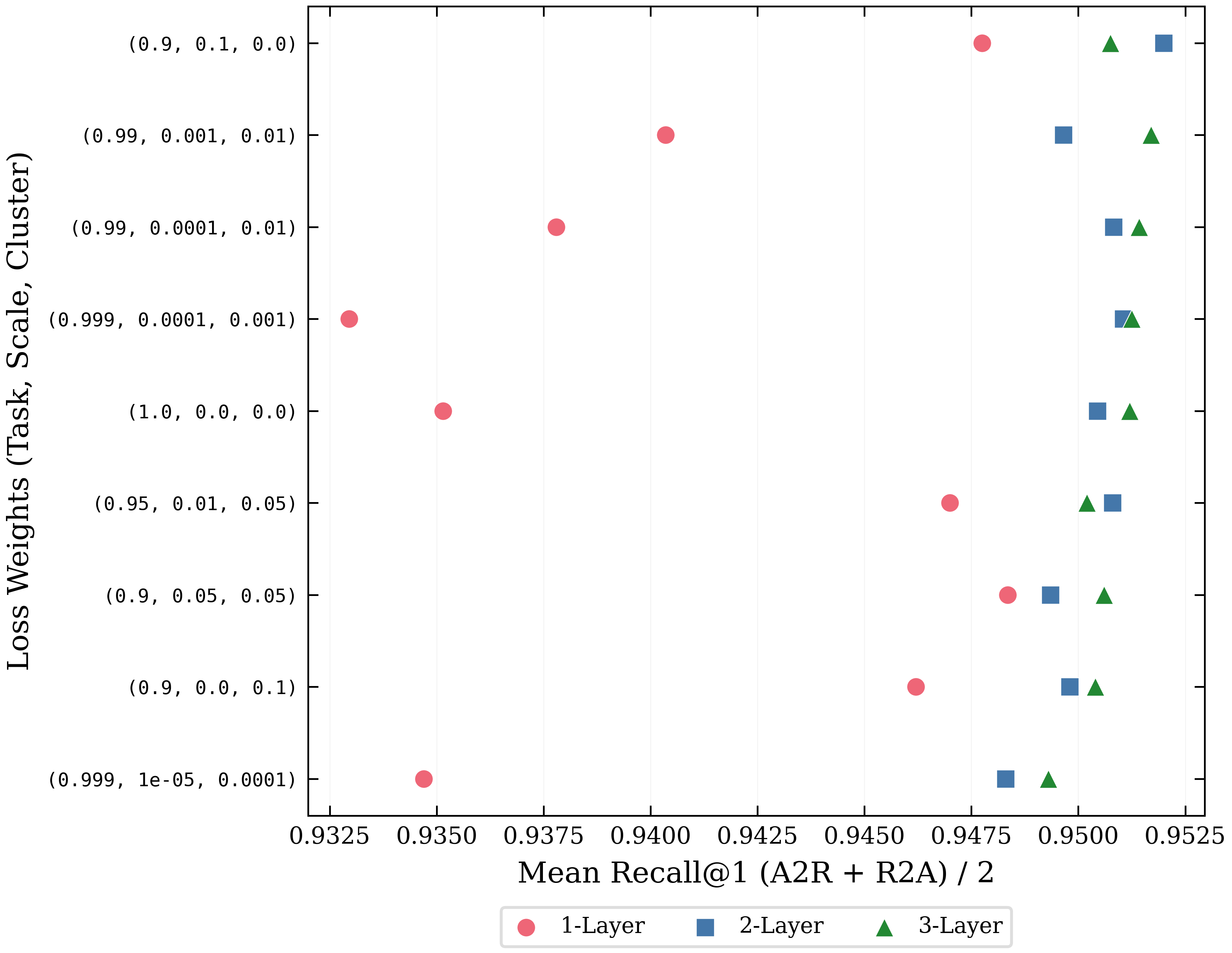}
    \caption{Mean recall (avg.\ of A2R and R2A) across loss weight configurations and network depths.}
    \label{fig:projection_ablation}
\end{figure}

Table~\ref{tab:projection_ablation} and Figure~\ref{fig:projection_ablation} summarize the paired-data recall performance of the adaptive projection network across 30 configurations varying in loss term weighting and network depth.

\begin{table}[H]
  \centering
  \caption{Projection network ablation: mean recall (avg.\ of A2R and R2A) across loss weight configurations and network depths. Bold indicates best per column.}
  \label{tab:projection_ablation}
  \small
  \setlength{\tabcolsep}{4pt}
  \begin{tabular}{ccccccc}
    \toprule
    \textbf{Task} & \textbf{Scale} & \textbf{Cluster} & \textbf{Layers} & \textbf{R@1} & \textbf{R@3} & \textbf{R@5} \\
    \midrule
    0.9 & 0.1 & 0 & 2 & $\bm{0.9520}$ & 0.9919 & 0.9972 \\
    0.99 & $1\text{e}-3$ & 0.01 & 3 & 0.9517 & 0.9920 & 0.9970 \\
    0.99 & $1\text{e}-4$ & 0.01 & 3 & 0.9514 & 0.9922 & 0.9972 \\
    0.999 & $1\text{e}-4$ & $1\text{e}-3$ & 3 & 0.9512 & 0.9919 & 0.9971 \\
    1 & 0 & 0 & 3 & 0.9512 & 0.9921 & 0.9973 \\
    0.999 & $1\text{e}-4$ & $1\text{e}-3$ & 2 & 0.9510 & 0.9919 & 0.9973 \\
    0.99 & $1\text{e}-4$ & 0.01 & 2 & 0.9508 & 0.9924 & 0.9970 \\
    0.95 & 0.01 & 0.05 & 2 & 0.9508 & 0.9921 & 0.9971 \\
    0.9 & 0.1 & 0 & 3 & 0.9507 & 0.9918 & 0.9972 \\
    0.9 & 0.05 & 0.05 & 3 & 0.9506 & 0.9917 & 0.9970 \\
    1 & 0 & 0 & 2 & 0.9505 & 0.9921 & 0.9971 \\
    0.9 & 0 & 0.1 & 3 & 0.9504 & 0.9921 & $\bm{0.9973}$ \\
    0.95 & 0.01 & 0.05 & 3 & 0.9502 & 0.9919 & 0.9970 \\
    0.9 & 0 & 0.1 & 2 & 0.9498 & 0.9920 & 0.9971 \\
    0.99 & $1\text{e}-3$ & 0.01 & 2 & 0.9496 & 0.9919 & 0.9969 \\
    0.9 & 0.05 & 0.05 & 2 & 0.9494 & 0.9921 & 0.9973 \\
    0.999 & $1\text{e}-5$ & $1\text{e}-4$ & 3 & 0.9493 & 0.9925 & 0.9972 \\
    0.9 & 0.05 & 0.05 & 1 & 0.9484 & 0.9919 & 0.9967 \\
    0.999 & $1\text{e}-5$ & $1\text{e}-4$ & 2 & 0.9483 & $\bm{0.9926}$ & 0.9972 \\
    0.9 & 0.1 & 0 & 1 & 0.9477 & 0.9922 & 0.9966 \\
    0.95 & 0.01 & 0.05 & 1 & 0.9470 & 0.9924 & 0.9968 \\
    0.9 & 0 & 0.1 & 1 & 0.9462 & 0.9916 & 0.9965 \\
    0.99 & $1\text{e}-3$ & 0.01 & 1 & 0.9404 & 0.9892 & 0.9962 \\
    0.99 & $1\text{e}-4$ & 0.01 & 1 & 0.9378 & 0.9888 & 0.9959 \\
    1 & 0 & 0 & 1 & 0.9351 & 0.9887 & 0.9955 \\
    0.999 & $1\text{e}-5$ & $1\text{e}-4$ & 1 & 0.9347 & 0.9879 & 0.9950 \\
    0.999 & $1\text{e}-4$ & $1\text{e}-3$ & 1 & 0.9329 & 0.9871 & 0.9946 \\
    \bottomrule
  \end{tabular}
\end{table}

Network depth has a pronounced effect on retrieval quality. 3-layer projections achieve the highest mean R@1 of $0.9508$, while 1-layer networks score $0.9408$---a relative gap of $1.07\%$. 

\begin{figure}[H]
    \centering
    \includegraphics[width=\linewidth]{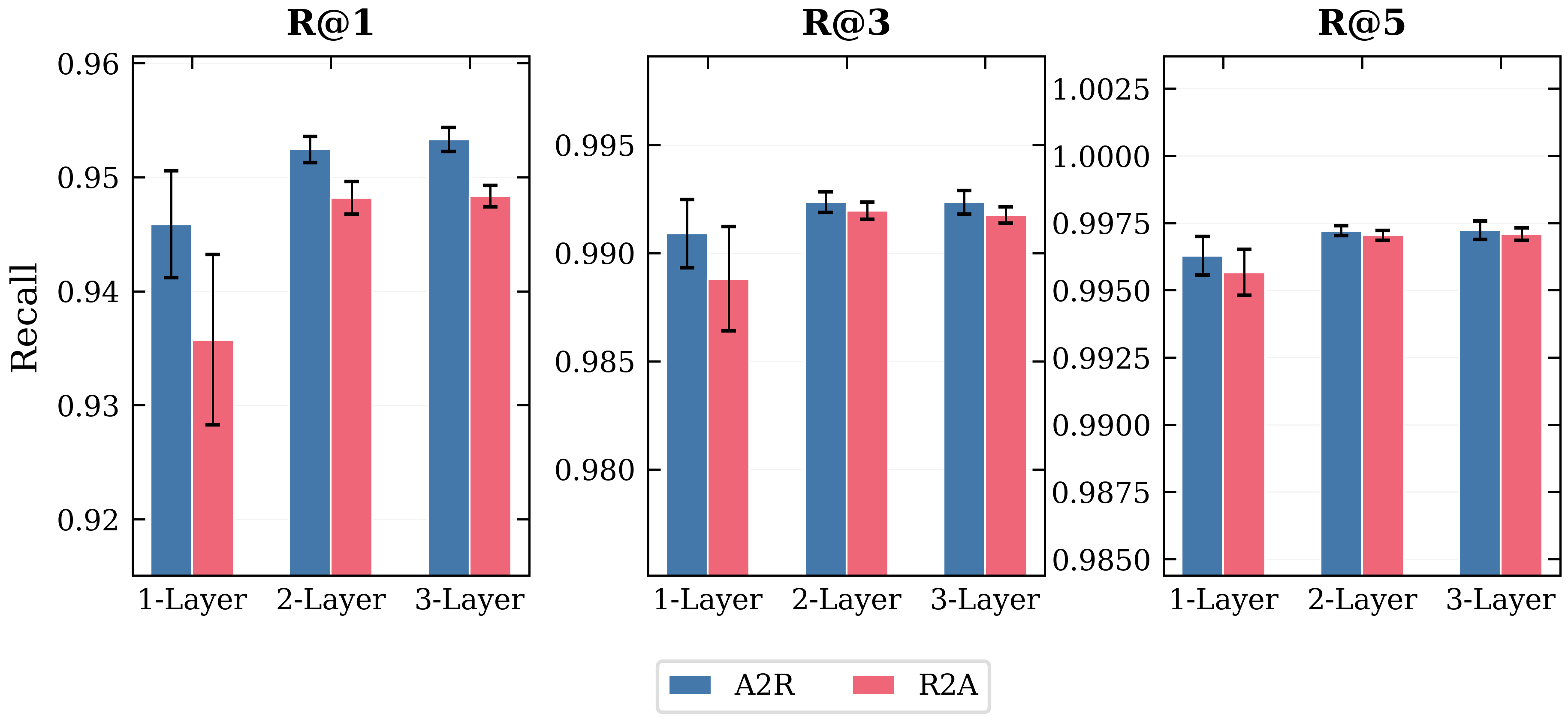}
    \caption{Projection layer depth effect on A2R \& R2A R@1,3,5.}
    \label{fig:projection_ablation}
\end{figure}

\begin{figure}[H]
    \centering
    \includegraphics[width=\linewidth]{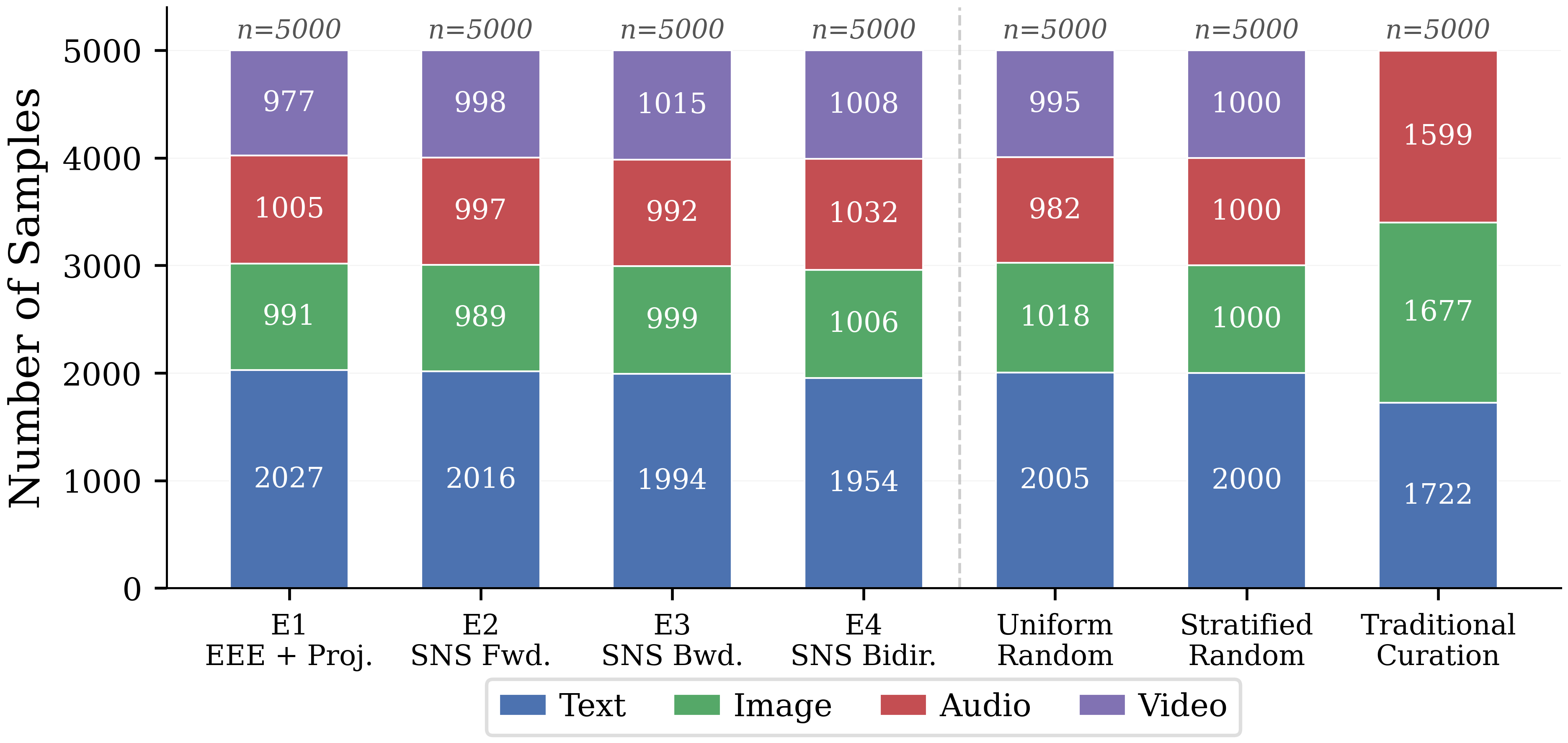}
    \caption{Datablend modality composition compared across downstream evaluation study curation strategies.}
    \label{fig:modality_dist}
\end{figure}

The effect of the projection network on modality composition of datablends is pronounced when comparing Figure \ref{fig:modality_dist} with Figure \ref{fig:datablend-modality-eee}. 

\clearpage

\subsubsection{Projection Network Embedding Space Geometry}
\label{ref:proj-geometry}

We also explore the embedding space geometry closer to understand the impact of the projection network on mitigating modality gap and non-semantic clustering tendencies of base embedding experts.

\begin{figure}[H]
    \centering
    % First subfigure
    \begin{subfigure}[t]{0.48\textwidth}
        \centering
        \includegraphics[width=\textwidth,height=0.3\textheight,keepaspectratio]{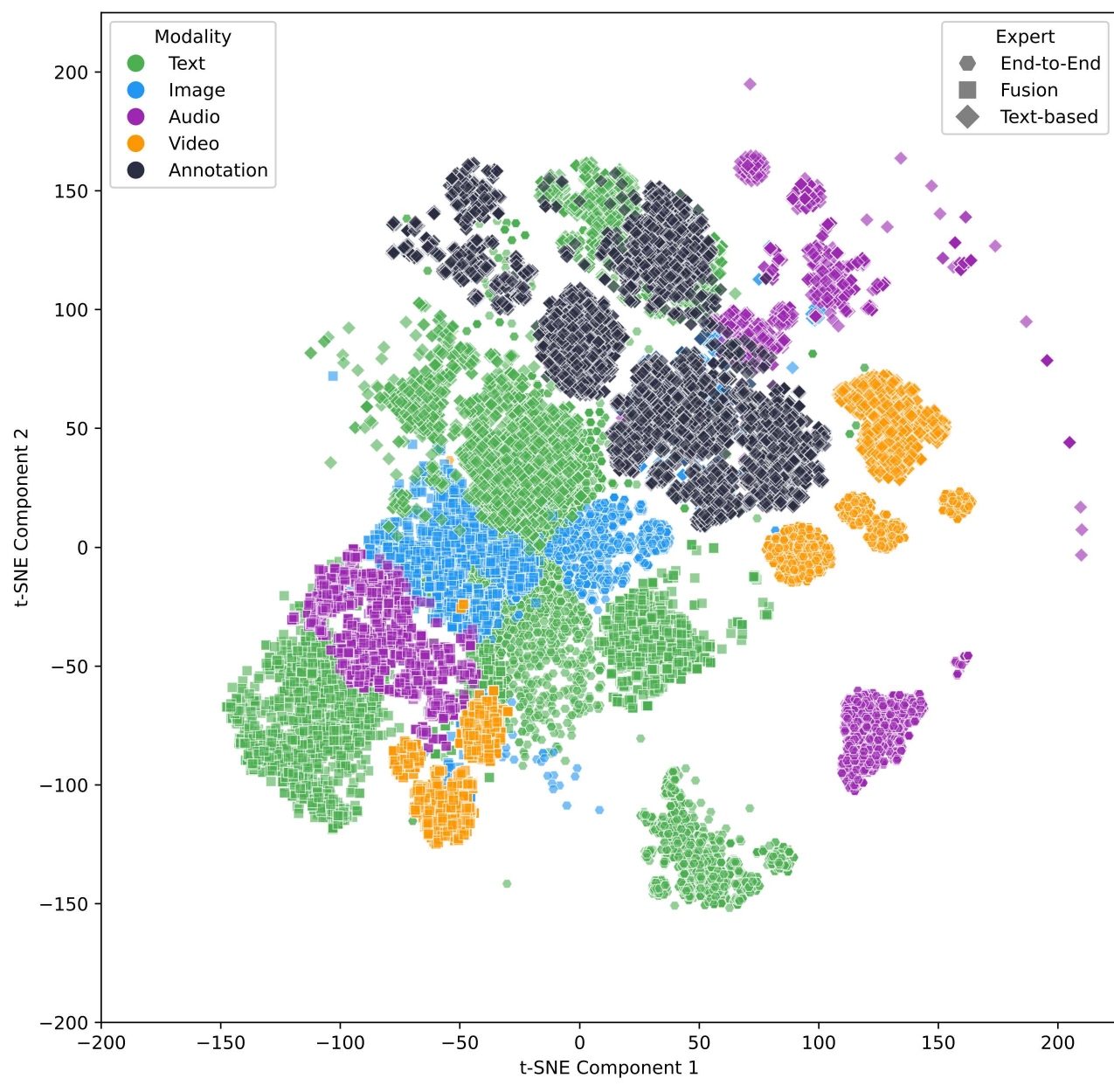}
        \caption{Isolated base expert embeddings (Fusion, Text-Based, End-to-End), no SNS.}
        \label{fig:before-proj}
    \end{subfigure}%
    \hfill
    % Second subfigure
    \begin{subfigure}[t]{0.48\textwidth}
        \centering
        \includegraphics[width=\textwidth,height=0.3\textheight,keepaspectratio]{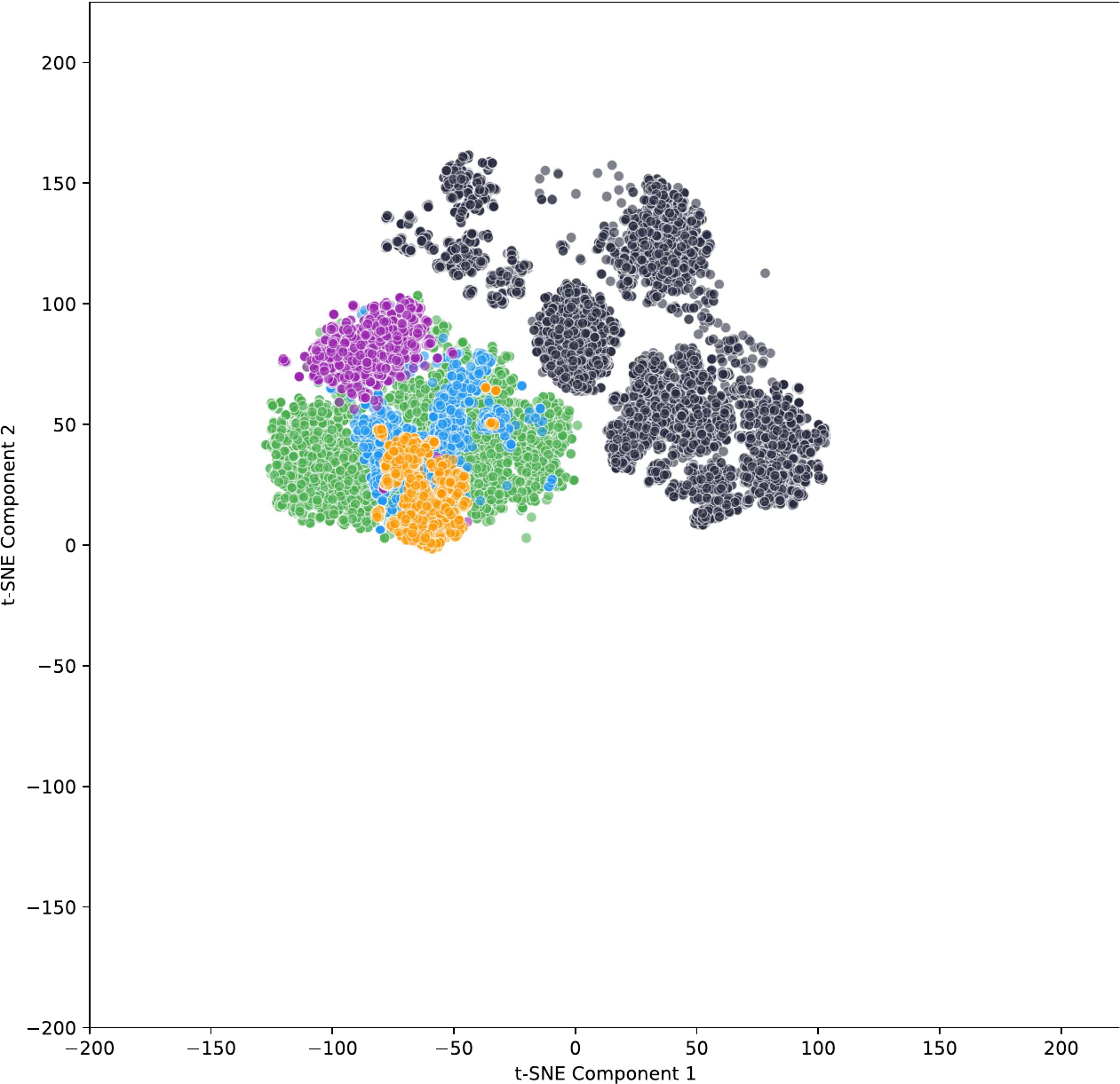}
        \caption{Fused embeddings after Projection Network}
        \label{fig:after-proj}
    \end{subfigure}

    \caption{2D t-SNE visualizations of embedding spaces pre- vs post- Projection Network. Modality gap clustering disappears after passing through the projection network. \textit{Note:} the grounded anchor embeddings for the text annotations are also displayed to show learned proximity between raw data embeddings $ \big[ e_{\text{fused}} \big]$ and the static annotation embeddings $\big[ a_{e_a} \big]$.}
    \label{fig:proj-pre-vs-post}
\end{figure}

\begin{table}[H]
\centering
\caption{Modality gap ($\ell_2$ embedding distance) across embedding spaces.}
\label{tab:modality-gaps}
\begin{tabular}{@{}lcccc|cccc@{}}
\toprule
 & \multicolumn{4}{c|}{\textbf{Gap}} & \multicolumn{4}{c}{\textbf{$\Delta$ vs.\ Projection (\%)}} \\
\textbf{Space} & \textbf{Video} & \textbf{Audio} & \textbf{Image} & \textbf{Text} & \textbf{Video} & \textbf{Audio} & \textbf{Image} & \textbf{Text} \\
\midrule
E2E Expert       & 45.73 & 45.39 & 29.09 & 19.50 & $-$99.8 & $-$99.8 & $-$99.7 & $-$99.6 \\
Fusion Expert    & 30.36 & 30.82 & 30.44 & 45.57 & $-$99.7 & $-$99.7 & $-$99.7 & $-$99.8 \\
Text Expert      &  0.441 &  0.294 &  0.265 &  0.257 & $-$82.1 & $-$67.3 & $-$67.4 & $-$72.4 \\
\textit{EEE + Projection (Ours)}       &  \textbf{0.079} &  \textbf{0.096} &  \textbf{0.086} &  \textbf{0.071} & --- & --- & --- & --- \\
\bottomrule
\end{tabular}
\end{table}

\clearpage

\subsection{Embedding Space Geometry: Base Multimodal Experts vs EEE + Projection}

\begin{figure}[H]
    \centering
    % First subfigure
    \begin{subfigure}[t]{0.48\textwidth}
        \centering
        \includegraphics[width=\textwidth,height=0.3\textheight,keepaspectratio]{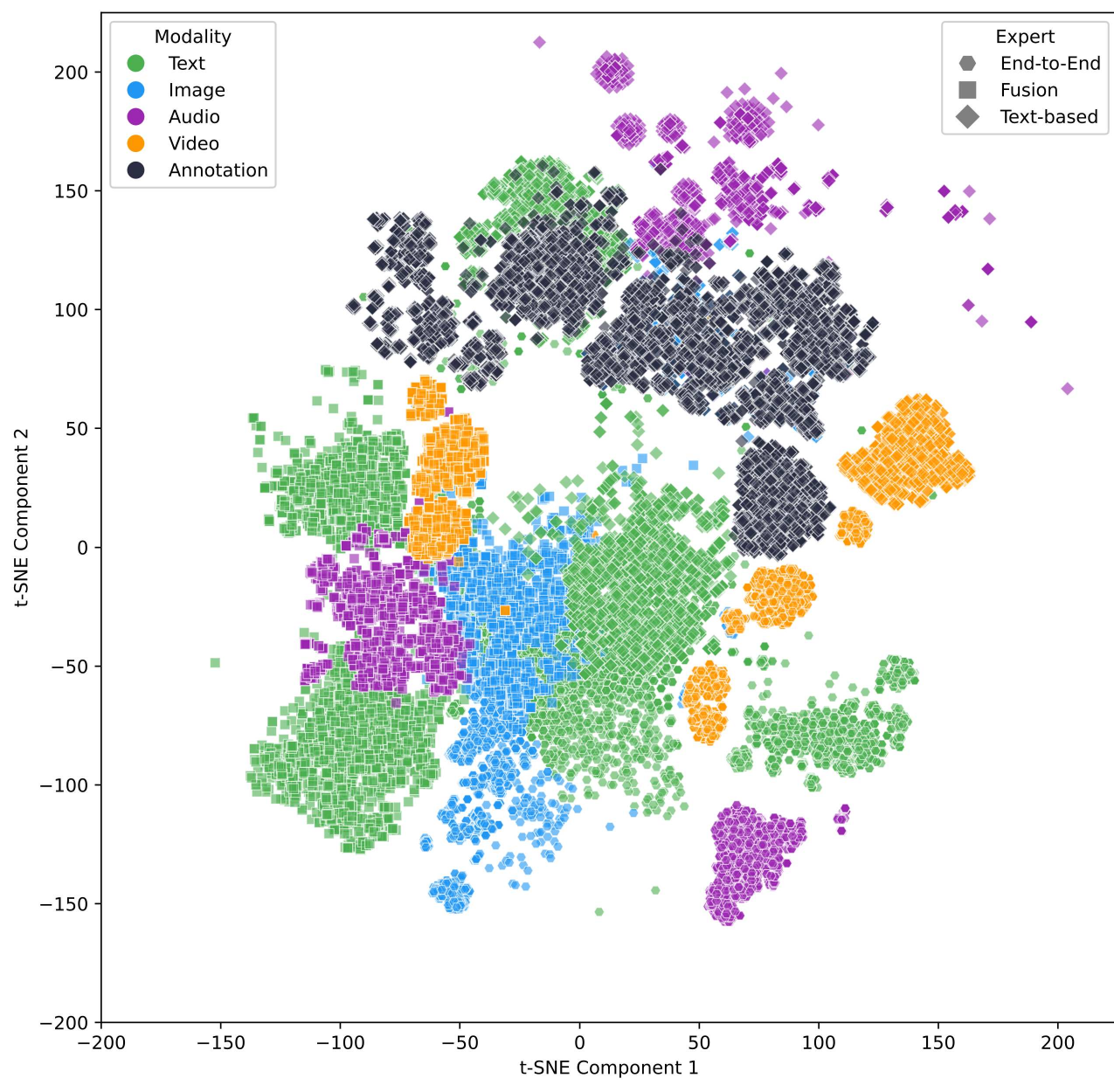}
        \caption{Base expert embeddings (Fusion, Text-Based, End-to-End)}
        \label{fig:before-proj}
    \end{subfigure}%
    \hfill
    % Second subfigure
    \begin{subfigure}[t]{0.48\textwidth}
        \centering
        \includegraphics[width=\textwidth,height=0.3\textheight,keepaspectratio]{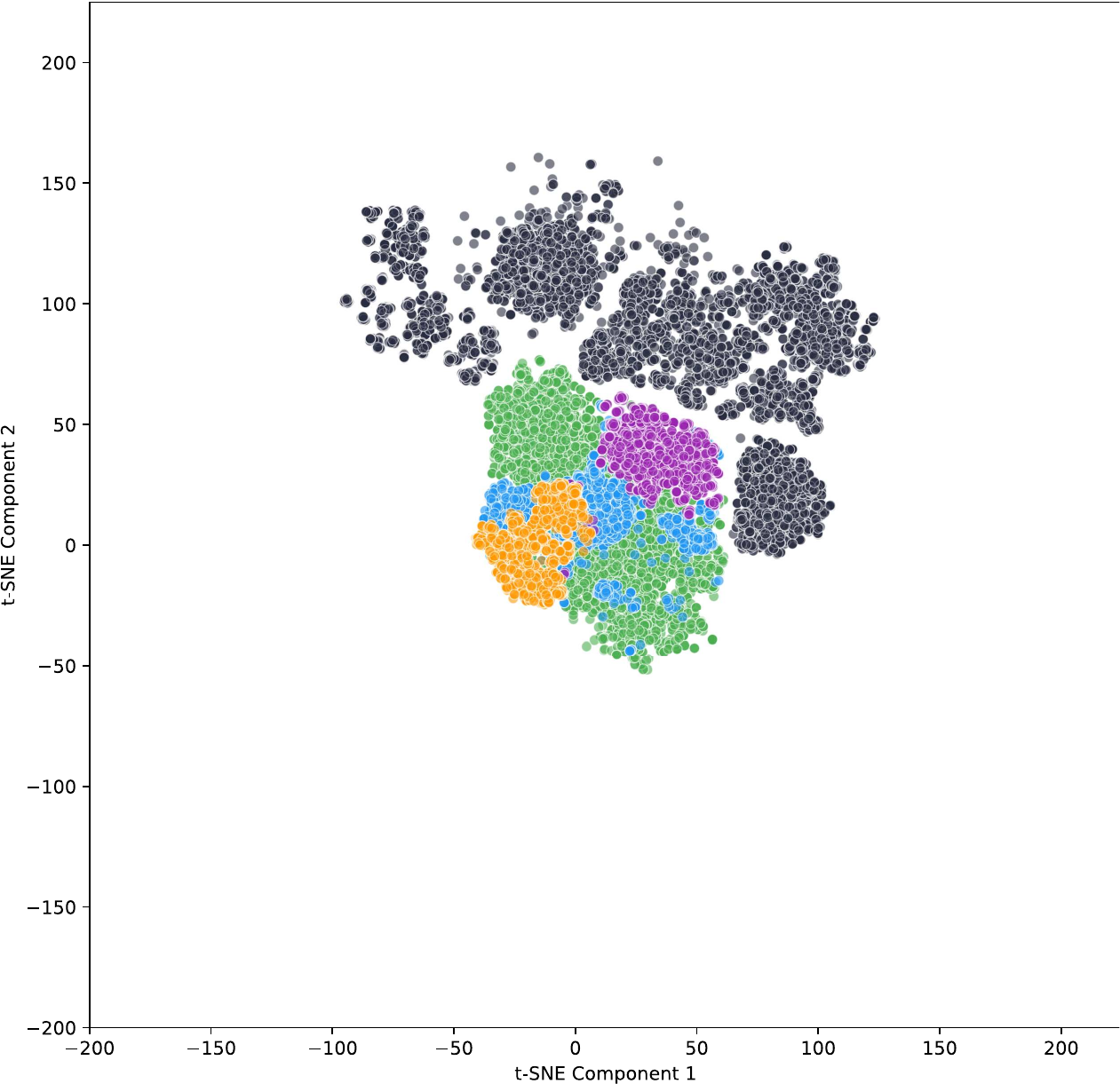}
        \caption{Fused embeddings after Projection Network}
        \label{fig:after-proj}
    \end{subfigure}

    \caption{2D t-SNE visualizations of embedding spaces without \textbf{(a).} and with \textbf{(b).} the projection network. Modality gap clustering is reduced by over 90\% on average vs base experts. All base experts and our approach here apply SNS pre-processing to samples prior to embedding. \textit{Note:} the grounded anchor embeddings for the text annotations are also displayed to show learned proximity between raw data embeddings $ \big[ e_{\text{fused}} \big]$ and the static annotation embeddings $\big[ a_{e_a} \big]$.}
    \label{fig:embed-space}
\end{figure}

\clearpage

\subsection{SNS Nucleus Examples}

Below, we show a couple of examples of multimodal nucleus extraction artifacts generated by the Symmetric Nucleus Subsampler (SNS) component, along with the $\Delta$ in approximate mutual information (MI).

\begin{figure}[H]
    \centering
    \includegraphics[width=\linewidth]{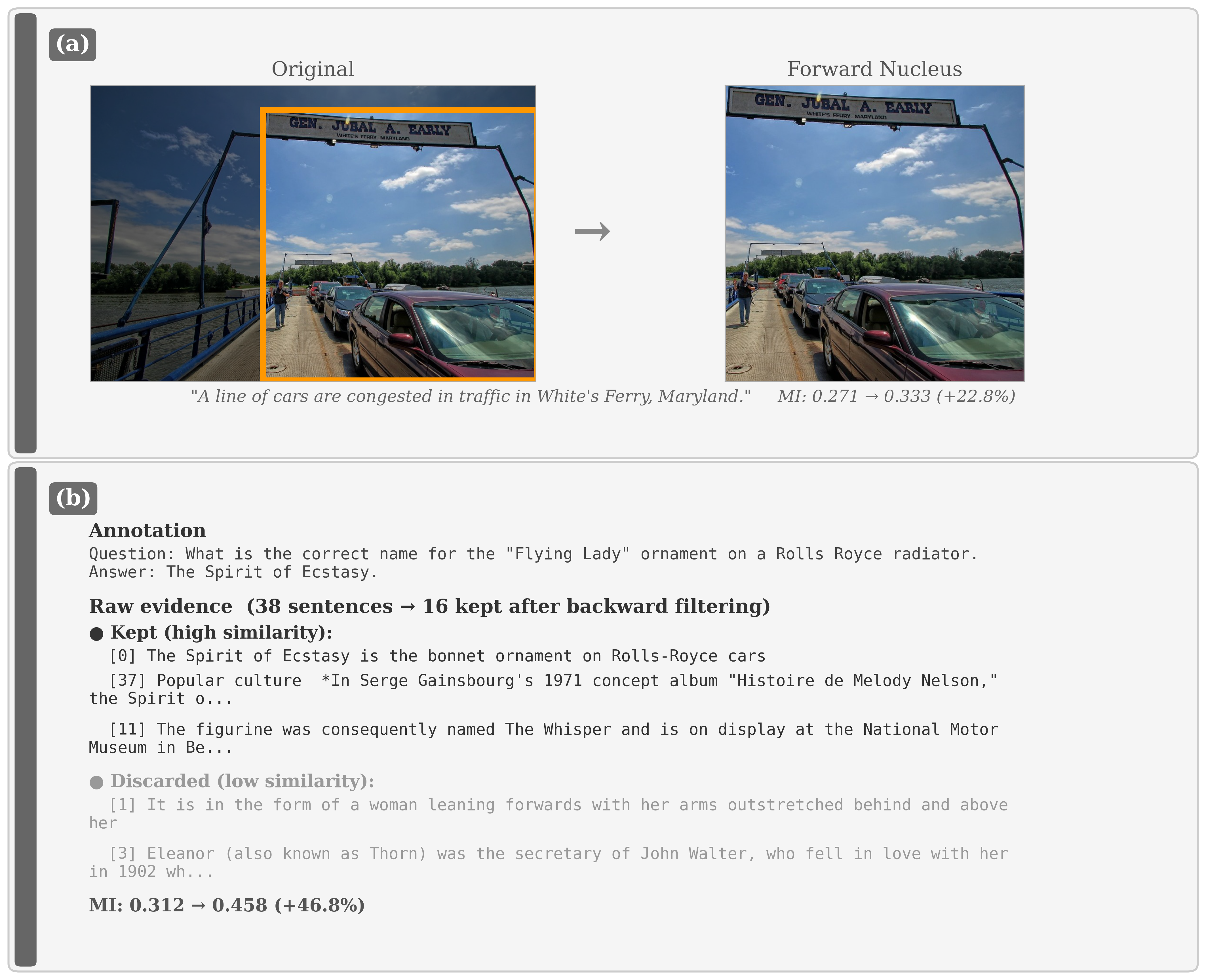}
    \caption{Pre- vs post-SNS examples. (a). Image sample from \textit{TextCaps},  (b). Text sample from \textit{TriviaQA}}
    \label{fig:sns_examples}
\end{figure}

\clearpage

\subsection{Downstream Evaluation Study}

\subsubsection{Baselines}
\label{ref:baselines}
\paragraph{``Traditional'' Data Curation Baseline}

We implement a baseline that combines heuristic filtering with semantic ranking on the annotations in our candidate pool of 10{,}000 samples across five data pools (ALFRED, AudioCaps, CoNaLa, TextCaps, TriviaQA). This pipeline utilizes NeMo Curator \cite{nemo-curator}, an open-source toolkit for scalable multimodal data curation, to filter and deduplicate samples, followed by embedding-based selection using a fixed pretrained NeMo retriever model (\texttt{llama-3.2-nv-embedqa-1b-v2}) \citep{nvidia2025llama32nvembedqa1bv2}. 

Unlike our proposed framework, which addresses raw-annotation misalignment jointly, this baseline operates in three decoupled stages: 

\begin{enumerate}
    \item \textbf{Unimodal Quality Filtering}: We first filter the annotations using text-based heuristic rules to remove low-quality supervision. Filters include a maximum non-alphanumeric character ratio ($\leq 0.45$), a maximum repeated line fraction ($\leq 0.7$), and boilerplate string removal. 
    \item \textbf{Semantic Deduplication}:  We perform exact and semantic deduplication within the annotation pool to prevent over-representation of common topics. We employ sentence-transformers/all-MiniLM-L6-v2 embeddings with K-means clustering ($k=100$) and apply a cosine similarity threshold ($\epsilon=0.05$) to prune redundant annotations and their paired raw sample.
    \item \textbf{Single-Encoder Ranking}: Surviving candidates are ranked by cosine similarity between their annotation embeddings and a fixed target query embedding (``natural, real-world scenes with objects, landscape, subjects, or people"). We select the top $k=5{,}000$ pairs for the final training mixture.
\end{enumerate}

This baseline represents a ``representation-level" selection strategy that assumes the encoder's geometry accurately reflects meaning, without explicitly correcting for modality bias or raw-annotation misalignment. 

\clearpage

\subsection{Additional related work}
\label{ref:related}

\subsubsection{Embedding-based dataset exploration and semantic search tools}
A complementary body of work uses pretrained embeddings to explore datasets through semantic search and interactive visualization. Spacewalker is closely aligned with our setting because it supports multimodal data including text, images, and video, and provides interactive traversal of representation spaces with configurable embedding backbones and projection methods for inspection and annotation \citep{heine2024spacewalker}.

\paragraph{Relation to our work.}
Spacewalker and related embedding-based exploration systems show that a shared representation space is a practical substrate for browsing neighborhoods, diagnosing clusters, and surfacing outliers across modalities. Our use of embeddings is different in purpose and scale. Rather than focusing on interactive inspection and manual labeling, we use embedding-based search programmatically to curate and merge many human-labeled multimodal datasets into a single training pool. We then apply SNS to denoise and reweight paired examples with respect to their labels or tags, and retrain an expert-based embedding engine on the curated mixture to improve retrieval and reduce modality gaps.

\subsubsection{Annotation aware subsampling}
Interpretability work such as Sufficient Input Subsets (SIS) identifies minimal subsets of an input that preserve a model's prediction, exposing spurious shortcuts and offering instance-level explanations \citep{carter2019sis}. While SIS is not a data selection method, it motivates the broader notion that subset structure can reveal which parts of an example carry signal versus noise.

\paragraph{Relation to our work.}
SNS is conceptually related to subset-based ideas but is designed for a different goal. SIS constructs sufficient subsets to explain individual model decisions, whereas SNS constructs nucleus variants for paired multimodal inputs and their labels or tags to improve the training distribution. SNS is also symmetric in that it gates informativeness in both directions, from raw modality to annotation and from annotation to raw modality. This symmetry is important when combining multiple labeled datasets with inconsistent taxonomies and variable annotation noise, and it supports our objective of improving retrieval behavior while reducing modality-driven separations in the learned embedding space.

\subsubsection{Additional embedding-based exploration tools}
Beyond Spacewalker \citep{heine2024spacewalker}, several systems instantiate embedding indexed exploration for specific data types. AEye focuses on scalable visualization and navigation for image datasets using embedding indexes and dimensionality reduction \citep{groetschla2024aeye}. For video collections, diveXplore provides embedding-based search and interactive exploration over videos using image and text embeddings \citep{leopold2025divexplore}. For text corpora, Texture combines structured attribute views with embedding-based overview and neighborhood search to support inspection of data quality and subset construction \citep{epperson2025texture}. These systems reinforce the utility of embedding spaces for navigation and inspection across modalities.

% \subsubsection{Data selection and curation benchmarks}
% Data selection and filtering are widely recognized as critical in large-scale representation learning, particularly for contrastive pretraining where web-scale corpora contain noise, duplication, and weakly aligned pairs. Recent work studies practical filtering strategies and training refinements that improve contrastive vision language pretraining under noisy data \citep{radenovic2023diht}. Dataset curation benchmarks and systematic comparisons of filtering, reweighting, and sampling strategies for training multimodal models highlight that quality and mixture choices can dominate downstream outcomes \citep{gadre2023datacomp}.

\end{document}

%% file: math_commands.tex
\usepackage{amsmath,amsfonts,bm}

\def\eqref#1{equation~\ref{#1}}
\def\1{\bm{1}}

\DeclareMathAlphabet{\mathsfit}{\encodingdefault}{\sfdefault}{m}{sl}
\SetMathAlphabet{\mathsfit}{bold}{\encodingdefault}{\sfdefault}{bx}{n}